\documentclass{statistical}
\usepackage{natbib}
\usepackage{hyperref}
\usepackage[T1]{fontenc}
\usepackage{ae,aecompl}
\usepackage{rotating}
\usepackage{color}
\usepackage{graphicx}
\usepackage{appendix}
\usepackage{longtable}
\usepackage{tikz}
\usetikzlibrary{positioning,arrows.meta,shapes.geometric}
\usepackage{orcidlink}

\usepackage{txfonts}

\begin{document} 

\titlerunning{Statistical hardness classification of ULXs}

\authorrunning{S. Allak et al.} 

 \title{A statistical hardness classification of ultraluminous X-ray sources using long-term \textit{Swift}/XRT monitoring}

 \titlerunning{Statistical hardness classification of ULXs}

 \author{Sinan Allak\orcidlink{0000-0001-7093-1079}\inst{1}\thanks{Corresponding author: 
\email{sinan.allak@uni-tuebingen.de}}, Wei Yu\orcidlink{0000-0002-3229-2453}\inst{1}, Menglei Zhou\orcidlink{0000-0001-8250-3338}\inst{1}, Honghui Liu\orcidlink{0000-0003-2845-1009}\inst{1} \and Andrea Santangelo\orcidlink{0000-0003-4187-9560}\inst{1}}

\institute{Institut für Astronomie und Astrophysik, Sand 1, 72076
Tübingen, Germany\\
\email{sinan.allak@uni-tuebingen.de}
}

\abstract{We present a homogeneous statistical framework to characterize observed long-term hardness behavior in 29 ultraluminous X-ray sources (ULXs) using over 3500 archival \textit{Swift} X-Ray Telescope (XRT) observations. Observations were classified as soft-band dominated, hard-band dominated, or uncertain using the hardness ratio and its confidence bounds. We define $HR$ as the ratio of the 1.5--10 to 0.3--1.5~keV count rates. We derived $f_{\rm soft}$ and $f_{\rm hard}$ and introduced the Crossing Index (CI), defined as the fraction of consecutive pairs of confidently classified observations that fall on opposite sides of the $HR=1$ boundary. These descriptors were analyzed with rule-based classification and unsupervised hierarchical clustering. The rule-based analysis identifies eight soft-band-dominated, five hard-band-dominated, and 16 mixed-band systems, while clustering reveals a similar three-group structure. Sources with known or candidate accretor types tend to occupy band-dominated groups, whereas unknown accretors are more common in the mixed-band group. All four sources with reported long-term X-ray modulations discussed in an orbital or super-orbital context belong to the rule-based mixed-band class. The three-cluster solution was preferred in 25 of 29 leave-one-out trials, while bootstrap tests found the band-dominated groups more stable than the mixed-band group. Because the classification uses observed count rates, these groups represent \textit{Swift}/XRT signatures rather than intrinsic spectral states. Response-folding tests show that the modeled hardness ratio depends on absorption column density and photon index. Bayesian reanalysis of a representative low-count source reduced CI from 56.3\% to 38.5\% but preserved its mixed-band classification. Random downsampling produced no systematic shift in CI; using half of the observations changed median CI by at most $\sim$10\% across four sources. Larger samples with independent absorption constraints can test links between these observational classes, accretor type, long-term periodic modulation, and other X-ray properties.}


 \maketitle

\section{Introduction}

ULXs are extragalactic, off-nuclear X-ray binaries with luminosities exceeding $\sim2\times10^{39}$ erg s$^{-1}$, surpassing the Eddington limit for conventional stellar-mass compact objects \citep{2017ARA&A..55..303K,2021AstBu..76....6F,2023NewAR..9601672K,2023arXiv230200006P}. Over the past decade, mounting observational evidence has established that the majority of ULXs are powered by super-Eddington accretion onto stellar-mass black holes or neutron stars, with their extreme luminosities arising from a combination of anisotropic emission, geometric beaming, and powerful radiation-driven outflows \citep{2007MNRAS.377.1187P,2009MNRAS.393L..41K,2016Natur.533...64P,2018MNRAS.479.3978K}.

A major breakthrough came with the detection of coherent X-ray pulsations from several ULXs, conclusively demonstrating that neutron stars can sustain accretion at highly super-Eddington rates \citep{2014Natur.514..202B,2014Natur.514..198M,2017Sci...355..817I,2018MNRAS.476L..45C,2018ApJ...863....9W,2020ApJ...895...60R,2025ApJ...994L..38D}. Cyclotron resonant scattering features (CRSFs) provide one of the few direct diagnostics of the magnetic-field strength near the compact object \citep{1992herm.book.....M,2012MmSAI..83..230C,2019A&A...622A..61S}. However, candidate CRSFs have been reported in only a handful of ULXs (e.g., \citealt{2018NatAs...2..312B, 2026A&A...710L..36A,2026A&A...707L..20C}), highlighting the need for indirect observational diagnostics that can be applied uniformly across larger ULX samples.

X-ray spectral variability provides one of the most powerful tools for probing the accretion physics of ULXs. Variations in spectral hardness reflect changes in the relative contributions of the accretion disk, the Comptonizing region, and radiatively driven winds, and are therefore closely linked to changes in the accretion geometry and mass accretion rate \citep{2013MNRAS.435.1758S,2015MNRAS.447.3243M}. Within the framework of super-Eddington accretion, spectral evolution is expected to depend on both the intrinsic accretion rate and the observer's viewing angle, as optically thick outflows can obscure the hard inner emission and produce characteristic transitions between hard and soft spectral states \citep{2009MNRAS.393L..41K,2018MNRAS.479.3978K}. Long-term hardness evolution provides valuable insight into the physical evolution of ULXs and offers a natural basis for statistically characterizing their spectral behavior.

Long-term monitoring with the \textit{Neil Gehrels Swift Observatory} X-Ray Telescope (\textit{Swift}/XRT) has opened a unique window into the temporal evolution of ULXs over timescales ranging from months to more than a decade. Unlike the deep pointed observations provided by \textit{Chandra} and \textit{XMM-Newton}, the extensive \textit{Swift}/XRT archive enables systematic investigations of spectral evolution across hundreds of epochs. Despite this wealth of long-term observations, previous studies have primarily focused on individual ULXs or small source samples, and homogeneous statistical comparisons of long-term hardness behavior across larger ULX samples remain limited. It therefore remains unclear whether the observed long-term hardness distributions contain systematic information that can be compared reproducibly across the ULX population. Developing such a framework provides a uniform observational description of long-term \textit{Swift}/XRT count-rate behavior, without assuming that the resulting groups correspond directly to intrinsic spectral states or compact-object types.

In this work, we present a homogeneous statistical characterization of the long-term observed \textit{Swift}/XRT hardness behavior of 29 ULXs. For each source, we measure the fractions of soft-band-dominated, hard-band-dominated, and uncertain observations and introduce CI as a dimensionless descriptor of the ordered hardness-label sequence. We then examine the structure of this descriptor space using a rule-based classification and complementary unsupervised hierarchical clustering. Finally, we compare the resulting observational groups with literature information on compact-object type and reported long-term modulation, while assessing the effects of absorption, low-count statistics, temporal sampling, and sample composition.

\section{Sample selection and observations}

\subsection{Sample selection}

The primary objective of this work is to investigate the long-term X-ray spectral evolution of ULXs using the extensive \textit{Swift}/XRT archive. Since the analysis is based on the statistical evolution of the hardness ratio, only sources satisfying specific observational requirements were considered. The final sample was selected according to three criteria: (i) the source must be spatially resolved with the angular resolution of \textit{Swift}/XRT, (ii) the source must have sufficient long-term \textit{Swift}/XRT monitoring to support a statistical hardness analysis, and (iii) the sample must include both ULXs with established or candidate compact-object classifications and ULXs whose compact objects remain unidentified, enabling comparisons between different compact-object populations.

First, only ULXs that could be spatially resolved with \textit{Swift}/XRT were included. The XRT point-spread function has a half-power diameter of approximately (18''), and sources located close to the host-galaxy nucleus or to nearby X-ray sources were excluded whenever their emission could not be spatially resolved. The adopted source coordinates were taken from previous high-resolution X-ray studies and verified using archival \textit{Chandra} observations whenever available.

Second, ULXs with at least 20 \textit{Swift}/XRT observations were preferentially selected to ensure sufficient temporal sampling for robust estimates of the statistical hardness fractions and the Crossing Index (CI). A small number of sources with fewer observations were nevertheless retained because they provide useful long-term light curves and broaden the comparison sample. Owing to the limited photon statistics of individual observations, hardness ratios could not be determined for every observation. In particular, observations with insufficient counts in either the soft or hard energy band were excluded from the hardness-ratio analysis, although they were retained in the long-term X-ray light curves. 

Third, the sample includes ULXs with confirmed or candidate compact-object classifications together with ULXs whose compact objects remain unidentified. This enables a direct comparison between systems with known compact-object classifications and those with unknown compact objects in order to investigate whether different compact-object populations exhibit distinct long-term statistical hardness behavior. The final sample consists of 29 sources. Table~\ref{tab:sample} summarizes the host galaxies, morphological types, adopted distances, and source coordinates of the selected sources.

\begin{table*}
\centering
\caption{ULX sample used in this work.}
\begin{tabular}{lccccccc}
\hline
No. & Galaxy & Distance (Mpc) & Morphology & Source & R.A. & Decl. & Class$^{a}$\\
\hline
1 & IC 10 & 0.7 & Irr & X--1 & 00:20:28.9 & +59:16:50.5 & BH-ULX$^{5}$ \\
2 & NGC 247 & 3.3 & SAB(s)d & ULX--1 & 00:47:03.7 & $-$20:47:45.6 \\
3 & NGC 300 & 2.0 & SA(s)d & X--1 & 00:55:09.7 & $-$37:42:10.8 & BH-ULX$^{7}$\\
4 & & & & ULX--1 & 00:55:04.86 & -37:41:43.7 & PULX$^{3}$ \\
5 & NGC 925 & 10.3 & SAB(rs)cd & ULX--2 & 02:27:21.52 & +33:35:00.8 \\
6 & & & & ULX--3 & 02:27:20.18 & +33:34:12.84 \\
7 & NGC 598 (M33) & 0.8 & SA(s)cd & X--7 & 01:33:34.1 & +30:32:07.3 \\
8 & NGC 1097 & 16.8 & SB(s)b & ULX--3 & 02:46:14.1 & $-$30:16:05.4 \\
9 & NGC 1313 & 4.6 & SB(s)d & X--2 & 3:18:22.21 & -66:36:04.33 & PULX$^{1}$ \\
10 & NGC 1566 & 17.7 & SAB(rs)bc & ULX--3 & 04:20:10.1 & $-$54:56:42.0 \\
11 & NGC 3521 & 11.5 & SAB(rs)bc & ULX--1 & 11:05:45.6 & +00:00:16.5 \\
12 & NGC 3623 & 12.6 & SSAB(rs)a & ULX--1 & 11:18:58.5 & +13:05:30.9 \\
13 & NGC 3627 & 11.1 & SSAB(s)b & ULX--2 & 11:20:15 & +12:59:30 \\
14 & NGC 3628 & 10.6 & Sb & ULX--1 & 11:20:15.72 & +13:35:14.10 \\
15 & NGC 4258 & 7.6 & SAB(s)bc & X--3 & 12:18:57.5 & +47:18:14.3 \\
16 & NGC 4490 & 7.8 & SB(s)d & X8 & 12:30:43.3 & +41:38:17.5 \\
17 & NGC 4559 & 7.3 & SAB(rs)cd & X--10 & 12:35:58.51 & +27:57:42.87 & \\
18 & & & & X--7 & 12:35:51.71 & +27:56:04.1 & PULX$^{10}$ \\
19 & NGC 5194 (M51) & 7.7 & SA(s)bc & ULX--2 & 13:29:43.5 & +47:11:35.23 \\
20 & & & & ULX--7 & 13:30:00.9 & +47:13:42.3 & PULX$^{6}$\\
21 & & & & ULX--8 & 13:30:07.6 & +47:11:06.7 & CRSF-ULX$^{7}$\\
22 & & & & ULX--eclipsing & 13:29:39.9 & +47:12:44.75 \\
23 & NGC 5236 (M83) & 4.7 & SAB(s)c & ULX--1 & 13:37:05.13 & -29:52:07.1 \\
24 & NGC 5643 & 16.9 & SAB(rs)c & X--1 & 14:32:42.15 & -44:09:36.10 \\
25 & NGC 5907 & 17.0 & SA(s)c & ULX--1 & 15:15:58.60 & +56:18:10.0 & PULX$^{8}$\\
26 & NGC 7456 & 15.7 & SA(rs)bc & ULX--1 & 23:02:05.62 & -39:36:17.0 & PULX$^{4}$ \\
27 & NGC 7793 & 3.4 & SA(s)d & P13 & 23:57:50.90 & -32:37:26.69 & PULX$^{2}$\\
28 & Milky Way & -- & -- & SS433 & 19:11:49.57 & +04:58:57.83 & \\
29 & Milky Way & -- & -- & Swift J0243.6+6124 & 02:43:40.43 & +61:26:03.76 &PULX$^{9}$ \\
\hline
\end{tabular}
\tablefoot{Compact-object identifications are adopted from the literature:
$^{1}$\citet{2019MNRAS.488L..35S}; $^{2}$\citet{2016ApJ...831L..14F}; $^{3}$\citet{2018MNRAS.476L..45C}; $^{4}$\citet{2026ApJ..1005...66Y}; $^{5}$\citet{2008ApJ...678L..17S}; $^{6}$\citet{2020ApJ...895...60R}; $^{7}$\citet{2019MNRAS.486....2M}; $^{8}$\citet{2010MNRAS.403L..41C}; \citet{2017Sci...355..817I}; $^{9}$\citet{2021MNRAS.500..565B}; $^{10}$\citet{2025A&A...695A.238P}.}
\label{tab:sample}
\end{table*}

\subsection{\textit{Swift}/XRT observations}

This work is based on archival observations obtained with the X-Ray Telescope (XRT; \citealp{2005SSRv..120..165B}) aboard the \textit{Neil Gehrels Swift Observatory} \citep{2004ApJ...611.1005G}. All publicly available photon counting (PC) mode observations were considered for each source in the final sample. Since several ULXs have been monitored over hundreds of individual observations, listing every observation separately is neither practical nor necessary. Instead, the observations were identified and retrieved through the UK Swift Science Data Centre (UKSSDC) using the \texttt{xrt\_prods} module of the \texttt{swifttools} Python package. This interface provides direct access to the UKSSDC online processing pipeline and ensures homogeneous retrieval of \textit{Swift}/XRT data products for the entire sample.

The number of observations varies from source to source, reflecting the heterogeneous long-term monitoring of individual ULXs within the \textit{Swift} archive. The complete data set comprises more than 3500 individual \textit{Swift}/XRT observations spanning nearly two decades. Owing to the limited photon statistics of some observations, reliable hardness ratios could not be determined for every observation, and only observations with reliable hardness-ratio measurements were included in the statistical hardness analysis. PC mode was adopted throughout this work because it provides two-dimensional imaging together with full spectral information, making it well suited for long-term studies of faint ULXs. Windowed Timing (WT) mode observations were excluded because they constitute only a small fraction of the available data and would introduce unnecessary heterogeneity into the statistical analysis.

\section{Data reduction and analysis}

All data used in this work were obtained from the public archive of the \textit{Swift}/XRT. The \textit{Swift}/XRT data products were generated using the \texttt{xrt\_prods} module of the \texttt{swifttools} package\footnote{https://www.swift.ac.uk/user\_objects/API/}, which accesses the online \textit{Swift}/XRT products generator maintained by the UK Swift Science Data Centre (UKSSDC; \citealt{2007A&A...469..379E,2009MNRAS.397.1177E}). The UKSSDC pipeline performs the standard \textit{Swift}/XRT calibration, screening, exposure correction, and background subtraction. All sources were processed using the same pipeline to ensure a homogeneous data reduction procedure. The count-rate products were generated with Bayesian bins enabled. At low count levels, the UKSSDC pipeline applies the Kraft--Burrows--Nousek method \citep{1991ApJ...374..344K} to estimate source count rates and their confidence bounds. The lower and upper count-rate uncertainties were retained separately throughout the hardness analysis. Measurements reported only as upper limits in either the soft or hard energy band were excluded from the hardness classification.

For each ULX, count rates were extracted in the total (0.3--10~keV), soft (0.3--1.5~keV), and hard (1.5--10~keV) energy bands. The adopted energy bands span the calibrated \textit{Swift}/XRT energy range. A boundary of 1.5~keV was adopted to separate the softer and harder spectral components while maintaining sufficient photon statistics in both energy bands for most observations. The hardness ratio was defined as \(HR = H/S\), where \(H\) and \(S\) denote the hard- and soft-band count rates, respectively. Reliable hardness ratios could not be obtained for all observations because of limited photon statistics in one or both energy bands. Therefore, the hardness--intensity, hardness--time, and Hard--Soft diagrams contain fewer data points than the corresponding long-term X-ray light curves.

Four diagnostic diagrams were generated for each source. Panel~(a) shows the long-term 0.3--10~keV light curve, where the observed count rates were converted to unabsorbed fluxes using PIMMS\footnote{\url{https://heasarc.gsfc.nasa.gov/cgi-bin/Tools/w3pimms/w3pimms.pl}} assuming an absorbed power-law spectrum with $\Gamma=1.7$ and a Galactic hydrogen column density of $N_{\rm H}=2\times10^{20}~{\rm cm^{-2}}$. A photon index of $\Gamma=1.7$ was adopted as a representative ULX continuum for the purpose of flux visualization only. Panels~(b), (c), and (d) present the hardness--intensity diagram (HID), the hardness--time diagram, and the Hard--Soft diagram, respectively. These diagnostic diagrams were designed to investigate complementary aspects of the long-term spectral evolution of ULXs. The long-term light curve characterizes the long-term X-ray variability of each source, allowing recurrent outbursts, bimodal flux distributions, long-term luminosity changes, and possible periodic or super-orbital modulations to be identified. The HID was constructed to investigate how the spectral hardness evolves as a function of X-ray luminosity. It allows us to examine whether individual ULXs maintain similar hardness levels or exhibit systematic hardness changes during their long-term evolution. Differences in the morphology of the HID may provide insight into changes in the accretion flow, the geometry of super-Eddington outflows, and the nature of the compact object.

The hardness--time diagram shows the chronological distribution of soft-band-dominated, hard-band-dominated, and uncertain observations over the full monitoring baseline. It also provides the ordered sequence of significant hardness labels used to calculate CI. Because CI contains no elapsed-time term, it describes the fraction of adjacent significant observations assigned to different hardness labels rather than the frequency or timescale of physical spectral transitions. Finally, the Hard--Soft diagram directly compares the hard- and soft-band count rates for every observation.

\begin{figure}
\resizebox{\hsize}{!}{\includegraphics{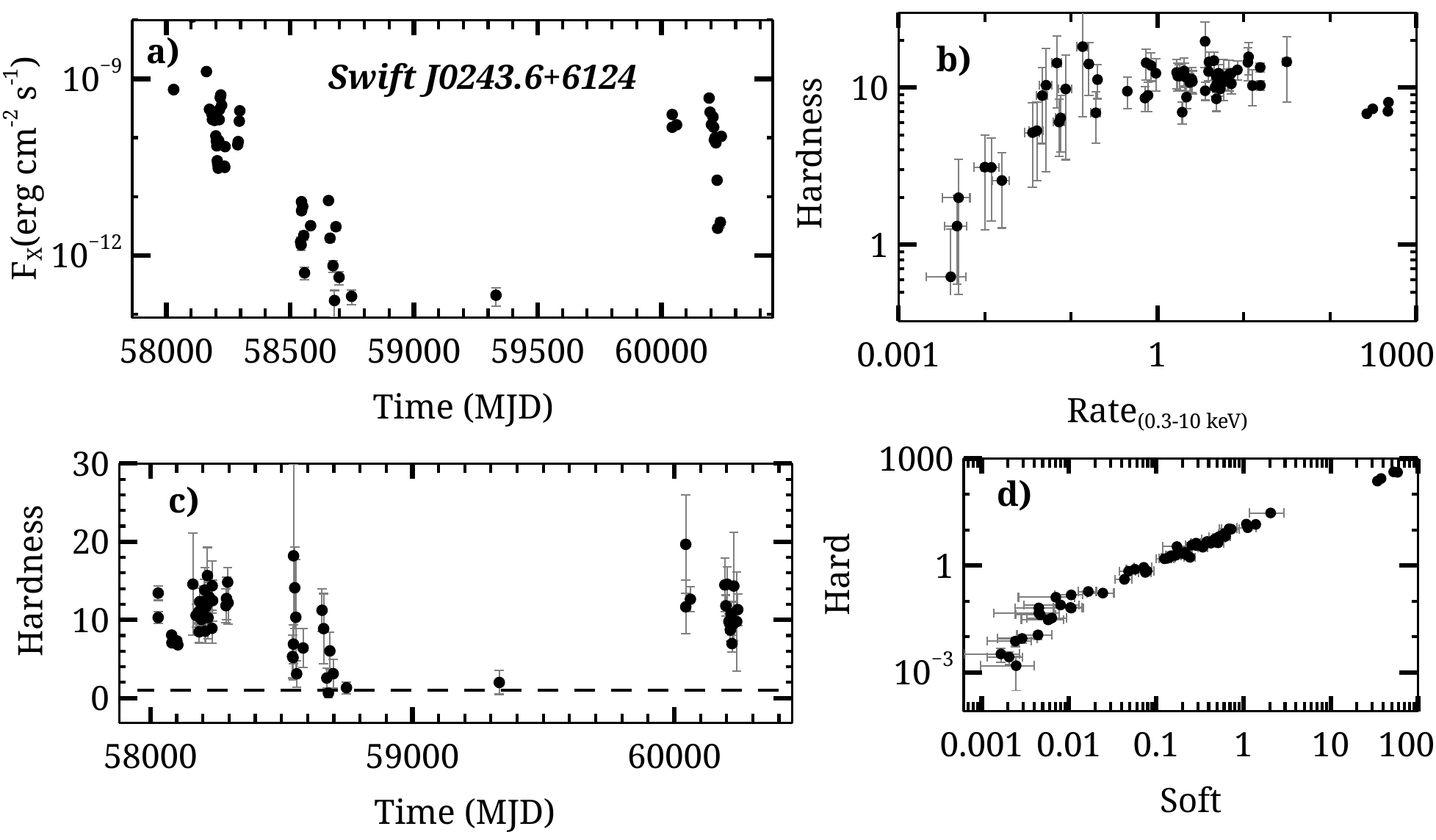}}
\caption{Representative long-term \textit{Swift}/XRT diagnostics for the Galactic ULX candidate \object{Swift J0243.6+6124}. Panels show (a) the 0.3--10~keV flux as a function of time (MJD), (b) the hardness ratio ($HR=H/S$) as a function of the total 0.3--10~keV count rate, (c) the hardness ratio as a function of time, and (d) the hard-band (1.5--10~keV) count rate versus the soft-band (0.3--1.5~keV) count rate. These four diagnostic diagrams constitute the basis of the statistical hardness analysis performed for all sources in this work.}
\label{fig:swiftj0243}
\end{figure}

\begin{figure}
\resizebox{\hsize}{!}{\includegraphics{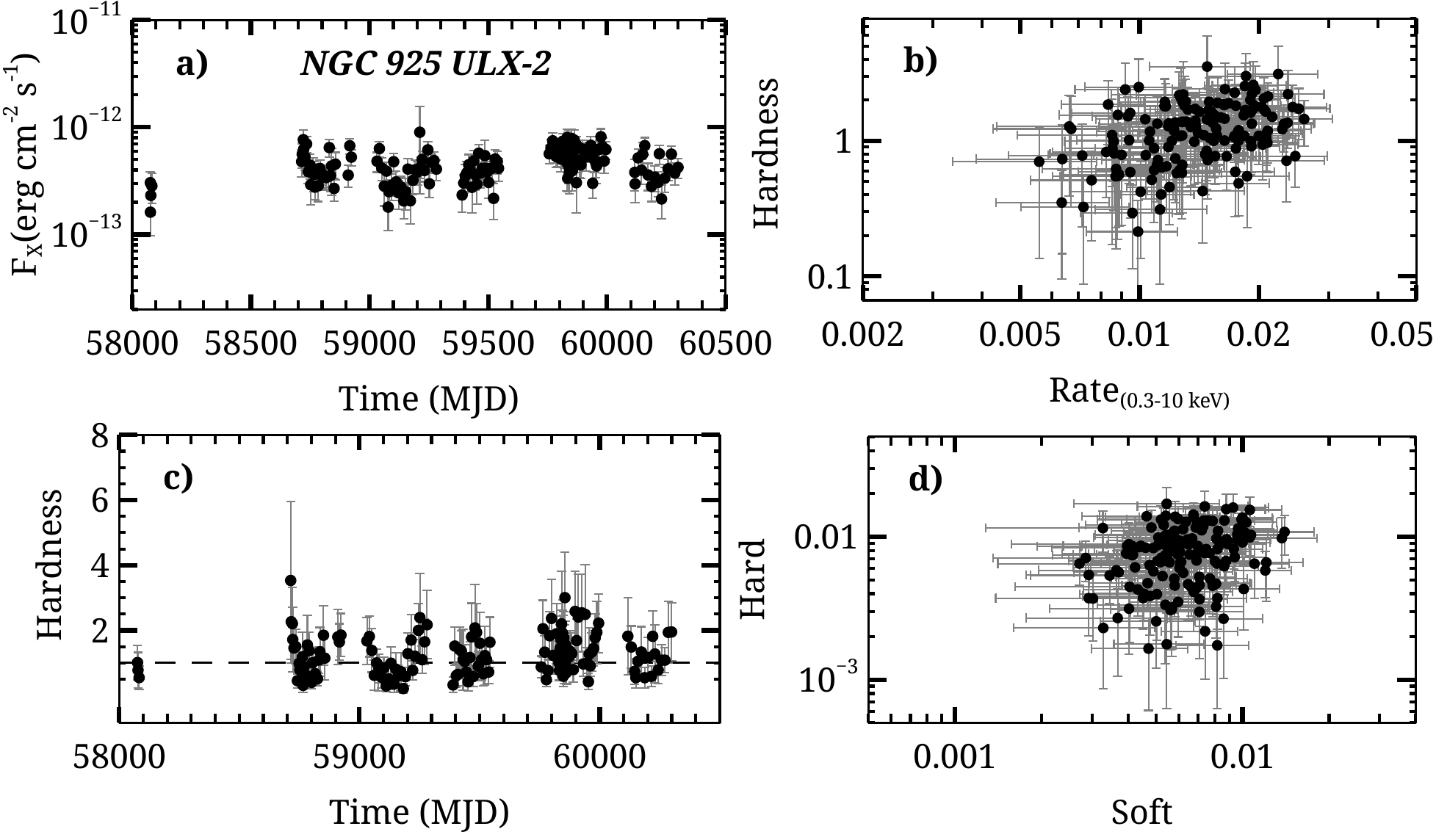}}
\caption{Same as Fig.~\ref{fig:swiftj0243}, but for NGC 925 ULX--2.}
\label{ngc925ulx2}
\end{figure}

\begin{figure}
\resizebox{\hsize}{!}{\includegraphics{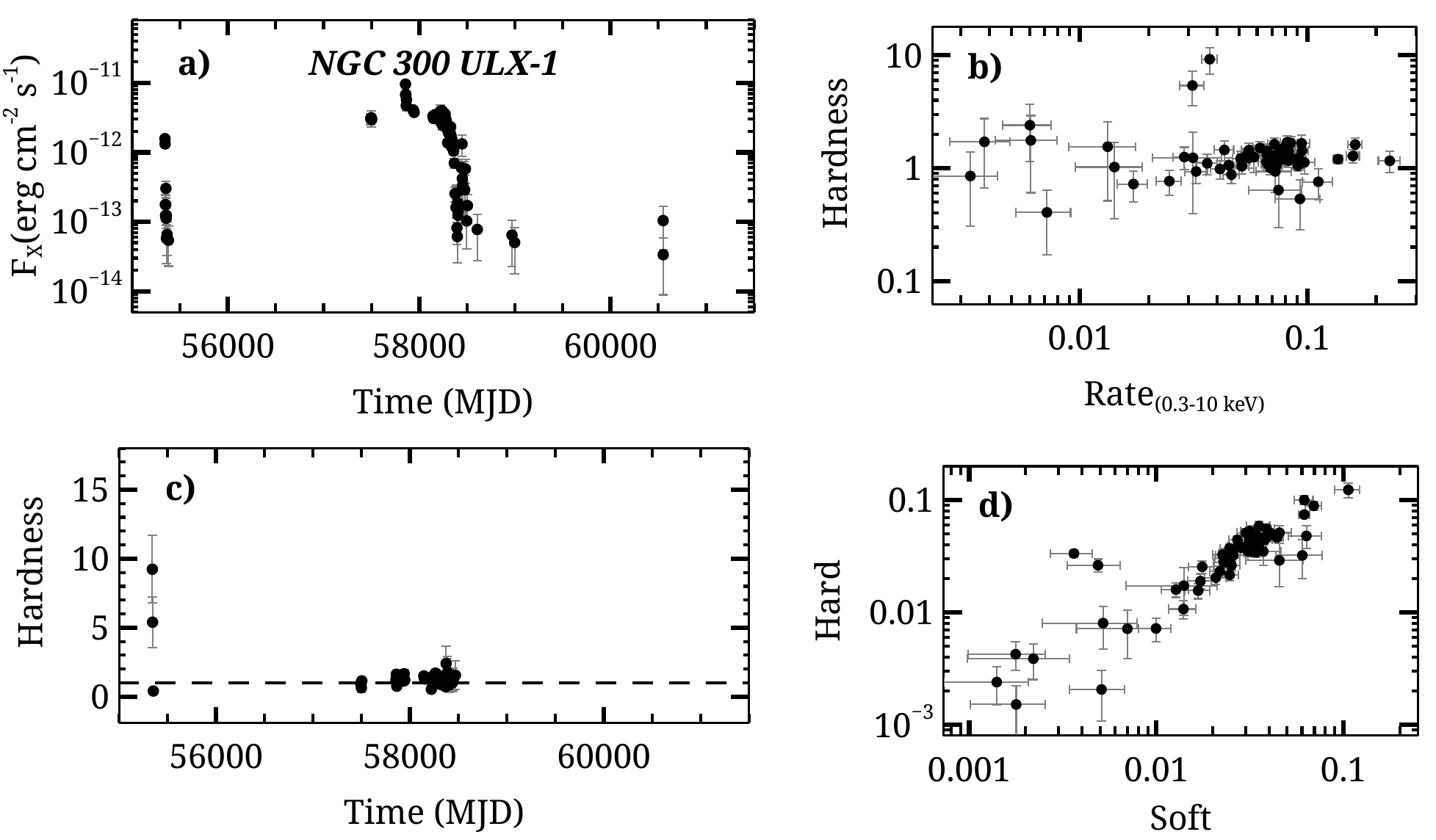}}
\caption{Same as Fig.~\ref{fig:swiftj0243}, but for NGC 300 ULX--1.}
\label{NGC300ULX1}
\end{figure}

Fluxes were used only for visualization of the long-term light curves, whereas all statistical analyses were performed using observed count rates. Examples of the four diagnostic diagrams are presented for the Galactic ULX pulsar Swift~J0243.6+6124, the ULX with an unknown compact object NGC~925 ULX--2, the pulsating ULX NGC~300 ULX--1, and the black-hole system IC~10 X-1 in Figs.~\ref{fig:swiftj0243}, \ref{ngc925ulx2}, \ref{NGC300ULX1}, and \ref{ic10X1}, respectively. The corresponding diagnostic diagrams for the remaining sources are presented in Appendix~\ref{figures in appen}. The long-term \textit{Swift}/XRT light curves were analyzed using the Lomb--Scargle periodogram \citep{1976Ap&SS..39..447L,1982ApJ...263..835S} to search for periodic variability over the full monitoring baseline. No previously unreported long-term periodicities with a false-alarm probability of \( {\rm FAP}<0.01 \) were detected. All significant periodic signals identified in the sample are consistent with previously reported periods in the literature.

\begin{figure}
\resizebox{\hsize}{!}{\includegraphics{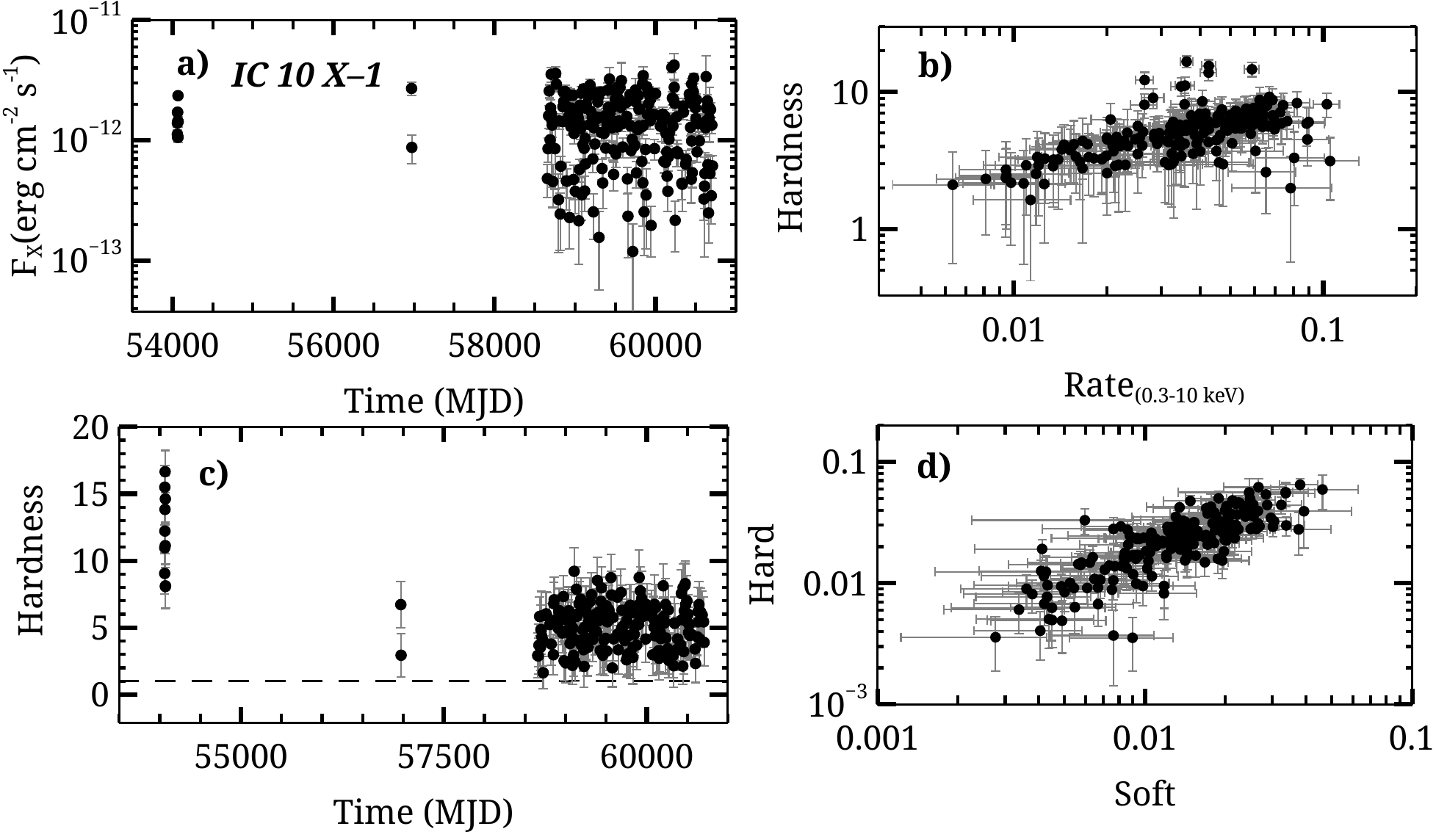}}
\caption{Same as Fig.~\ref{fig:swiftj0243}, but for IC 10 X-1.}
\label{ic10X1}
\end{figure}

\subsection{Statistical classification of the long-term hardness evolution} \label{sec:hardness_classification} 

To characterize the long-term observed hardness of each source, every \textit{Swift}/XRT observation with valid count-rate measurements in both energy bands was classified using its hardness ratio and the corresponding confidence bounds. The hardness ratio was defined as $HR=H/S$, where $H$ and $S$ are the hard- and soft-band count rates, respectively. Because the count-rate uncertainties may be asymmetric, their lower and upper values were retained separately. For each observation, we defined $H_{\rm lo}=H-\Delta H_{-}$, $H_{\rm hi}=H+\Delta H_{+}$, $S_{\rm lo}=S-\Delta S_{-}$, and $S_{\rm hi}=S+\Delta S_{+}$. Conservative hardness bounds were then calculated as $HR_{\rm lo}=H_{\rm lo}/S_{\rm hi}$ and $HR_{\rm hi}=H_{\rm hi}/S_{\rm lo}$.

The boundary $HR=1$ corresponds to equal observed count rates in the two energy bands. An observation was classified as soft-band dominated when $HR_{\rm hi}<1$ and as hard-band dominated when $HR_{\rm lo}>1$. Observations satisfying $HR_{\rm lo}\leq1\leq HR_{\rm hi}$ were classified as uncertain. This conservative procedure assigns an observation to a hardness regime only when its full confidence interval lies on one side of the adopted boundary. Measurements reported only as upper limits in either energy band were excluded from the hardness classification. For each source, the fractions of soft-band-dominated, hard-band-dominated, and uncertain observations were calculated as $f_{\rm soft}=N_{\rm soft}/N_{\rm obs}$, $f_{\rm hard}=N_{\rm hard}/N_{\rm obs}$, and $f_{\rm unc}=N_{\rm unc}/N_{\rm obs}$, where $N_{\rm obs}$ denotes the number of observations included in the hardness analysis.

To characterize changes in the ordered hardness-label sequence, uncertain observations were removed, and the remaining significant observations were ordered chronologically. A crossing was recorded whenever two adjacent observations in this reduced sequence were located on opposite sides of $HR=1$. The Crossing Index was defined as $CI=N_{\rm tr}/(N_{\rm sig}-1)$, where $N_{\rm tr}$ is the number of observed crossings and $N_{\rm sig}=N_{\rm soft}+N_{\rm hard}$. CI therefore measures the fraction of adjacent significant observations assigned to different hardness labels. It contains no elapsed-time term and does not measure a transition rate or physical variability timescale. If uncertain observations occur between two significant observations, those significant observations remain adjacent in the reduced sequence; consequently, the observed crossing does not constrain the source behavior during the intervening interval. CI is therefore conditional on the available \textit{Swift}/XRT sampling.

Finally, sources with significant observations only below $HR=1$ were classified as soft-band dominated, while those with significant observations only above $HR=1$ were classified as hard-band dominated. Sources with significant observations on both sides of the boundary were classified as mixed-band systems. These classes describe observed \textit{Swift}/XRT count-rate signatures and should not be interpreted as absorption-corrected ULX spectral states.

\begin{table*}
\caption{Statistical hardness properties, rule-based classification, and clustering results for the ULX sample.}
\label{tab:hardness_classification}
\centering
\begin{tabular}{lcccccccc}
\hline\hline
Source & $N_{\rm obs}$ & Soft (\%) & Hard (\%) & Uncertain (\%) & $N_{\rm tr}$ & CI (\%) & Class$^{a}$ & Cluster ID$^{b}$ \\
\hline
IC 10 X--1 & 204 & 0 & 95 & 5 & 0 & 0 & hard-band dominated & 2 \\
M33 X--7 & 18 & 39 & 11 & 50 & 1 & 12.5 & Mixed hardness & 1\\
M51 ULX--2 & 59 & 75 & 0 & 25 & 0 & 0 & soft-band dominated & 3\\
M51 ULX--7 & 216 & 39 & 6 & 56 & 22 & 23 & Mixed hardness & 1\\
M51 ULX--8 & 250 & 31 & 8 & 61 & 30 & 31 & Mixed hardness & 1\\
M51 ULX--eclipsing & 63 & 21 & 2 & 78 & 2 & 15 & Mixed hardness & 1\\
M83 ULX--1 & 63 & 57 & 2 & 41 & 2 & 5 & Mixed hardness & 3\\
NGC 247 ULX--1 & 41 & 100 & 0 & 0 & 0 & 0 & soft-band dominated & 3\\
NGC 300 X--1 & 180 & 92 & 0 & 8.3 & 0 & 0 & soft-band dominated & 3\\
NGC 300 ULX--1 & 67 & 9 & 52 & 39 & 8 & 20 & Mixed hardness & 2 \\
NGC 925 ULX--2 & 154 & 16 & 18 & 67 & 16 & 32 & Mixed hardness & 1\\
NGC 925 ULX--3 & 53 & 21 & 11 & 68 & 9 & 56 & Mixed hardness & 1\\
NGC 1313 X--2 & 507 & 14 & 38 & 48 & 64 & 25 & Mixed hardness & 2\\
NGC 1097 ULX--3 & 26 & 46 & 8 & 46 & 4 & 31 & Mixed hardness & 1\\
NGC 1566 ULX--3 & 31 & 42 & 0 & 58 & 0 & 0 & soft-band dominated & 3\\
NGC 3521 ULX--1 & 8 & 75 & 0 & 25 & 0 & 0 & soft-band dominated & 3\\
NGC 3623 ULX--1 & 9 & 44 & 0 & 56 & 0 & 0 & soft-band dominated & 3\\
NGC 3627 ULX--2 & 34 & 0 & 44 & 56 & 0 & 0 & hard-band dominated & 2\\
NGC 3628 ULX--1 & 39 & 2.6 & 49.7 & 49.7 & 2 & 11 & Mixed hardness & 2\\
NGC 4258 X--3 & 25 & 12 & 32 & 56 & 5 & 50 & Mixed hardness & 1\\
NGC 4490 X8 & 36 & 0 & 22 & 78 & 0 & 0 & hard-band dominated & 2\\
NGC 4559 X--10 & 129 & 31 & 7 & 62 & 12 & 25 & Mixed hardness & 1\\
NGC 4559 X--7 & 183 & 82 & 0 & 18 & 0 & 0 & soft-band dominated & 3\\
NGC 5643 X--1 & 89 & 16 & 13 & 71 & 10 & 40 & Mixed hardness & 1\\
NGC 5907 ULX--1 & 269 & 1 & 72 & 27 & 4 & 2 & Mixed hardness & 2\\
NGC 7456 ULX--1 & 37 & 86 & 0 & 14 & 0 & 0 & soft-band dominated & 3\\
NGC 7793 P13 & 407 & 2 & 71 & 27 & 12 &4 & Mixed hardness & 2\\
SS433 & 27 & 0 & 100 & 0 & 0 & 0 & hard-band dominated & 2\\
Swift J0243.6+6124 & 64 & 0 & 95 & 5 & 0 & 0 & hard-band dominated & 2\\
\hline
\end{tabular}
\tablefoot{$^{a}$The rule-based classification is based on the statistical hardness properties defined in Sec.~\ref{sec:hardness_classification}, whereas the cluster IDs$^{b}$ are obtained from the unsupervised hierarchical clustering analysis (Sec.~\ref{sec:makine}). Cluster~1 corresponds to the mixed-band group, Cluster~2 to the hard-band-dominated group, and Cluster~3 to the soft-band-dominated group.}
\end{table*}

\subsection{Unsupervised classification of statistical hardness properties} \label{sec:makine}

The rule-based classification described in Sect.~\ref{sec:hardness_classification} assigns source-level labels according to the distribution of significant observations around $HR=1$. As a complementary analysis, we applied unsupervised hierarchical clustering to $f_{\rm soft}$, $f_{\rm hard}$, and CI. No rule-based labels, decision boundaries, or compact-object information were supplied to the clustering algorithm. However, because the clustering uses the same descriptors that underlie the rule-based classification, it is not an independent physical validation. Instead, it tests whether these descriptors form a coherent statistical structure within the adopted observed-count parameter space.

Before clustering, the three input features were standardized using the z-score transformation $z_i=(x_i-\mu)/\sigma$, where $\mu$ and $\sigma$ denote the sample mean and standard deviation of each feature, respectively. Hierarchical agglomerative clustering was then performed using Ward linkage with the Euclidean distance metric, as implemented in the \texttt{scikit-learn} Python package \citep{Ward01031963, JMLR:v12:pedregosa11a}. The optimal number of clusters was determined by comparing solutions with $k=2$--6 using the silhouette coefficient \citep{ROUSSEEUW198753} as the primary evaluation metric because it simultaneously measures cluster compactness and separation. The Calinski--Harabasz \citep{Caliski01011974} and Davies--Bouldin \citep{4766909} indices were computed as complementary validation metrics.

\begin{figure}
\resizebox{\hsize}{!}{\includegraphics{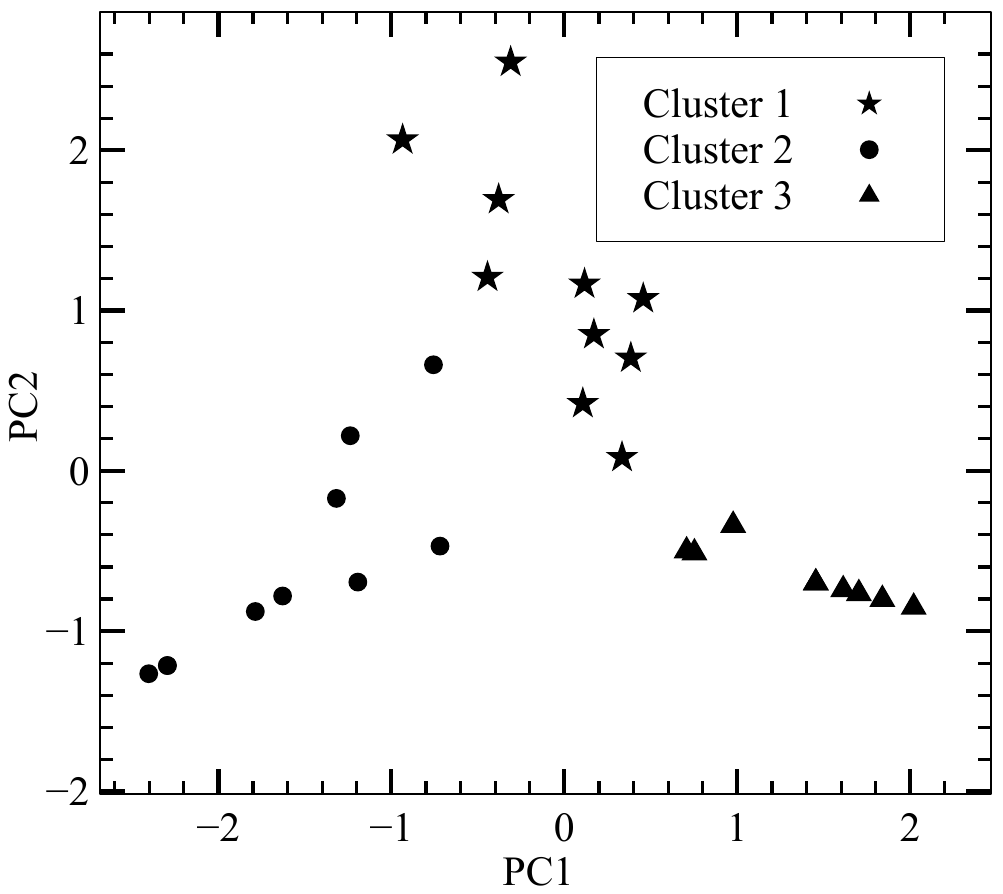}}
\caption{Principal Component Analysis (PCA) projection of the standardized statistical hardness parameter space defined by \(f_{\rm soft}\), \(f_{\rm hard}\), and the CI. Stars, circles, and triangles represent Clusters 1, 2, and 3, respectively, as identified by the unsupervised hierarchical clustering analysis. Hierarchical agglomerative clustering with Ward linkage was performed in the original three-dimensional feature space, while the PCA projection is shown solely for visualization. The first two principal components account for 94.5\% of the total variance.}
\label{fig:pca_clusters}
\end{figure}

\begin{figure}
\resizebox{\hsize}{!}{\includegraphics{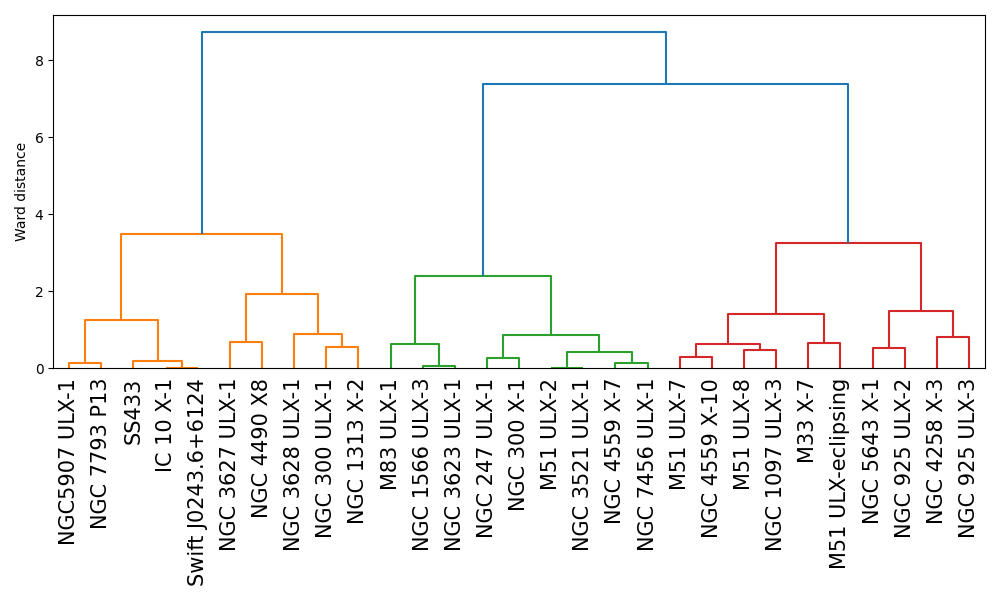}}
\caption{Hierarchical clustering dendrogram of the standardized statistical hardness descriptors \(f_{\rm soft}\), \(f_{\rm hard}\), and \(CI\), obtained using Ward linkage. The dashed horizontal line indicates the cut adopted for the three-cluster solution selected by the silhouette criterion.}
\label{fig:dendrogram}
\end{figure}

\begin{figure*}
\centering
\resizebox{0.82\textwidth}{!}{
\begin{tikzpicture}[
node distance=4mm and 9mm,
box/.style={
draw,
rounded corners,
align=center,
minimum width=4.5cm,
minimum height=6mm,
font=\scriptsize
},
arrow/.style={-{Latex[length=1.8mm]},thick}
]

\node[box] (a) {\textit{Swift}/XRT observations};
\node[box,below=of a] (b) {$HR=H/S$};

\node[box,below=of b] (c) {Observation hardness label\\
Soft-band dominated: $HR_{\rm hi}<1$\\
Uncertain: $HR_{\rm lo}\leq1\leq HR_{\rm hi}$\\
Hard-band dominated: $HR_{\rm lo}>1$};

\node[box,below=of c] (d) {Hardness descriptors\\
$f_{\rm soft}$,\ $f_{\rm hard}$,\ $CI$};

\node[box,below left=8mm and 10mm of d] (e1)
{Rule-based classification\\
soft-band / mixed-band / hard-band};

\node[box,below right=8mm and 10mm of d] (e2)
{z-score standardization};

\node[box,below=of e2] (f2)
{Ward hierarchical clustering};

\node[box,below=of f2] (g2)
{Cluster selection\\
Silhouette, CH, DB};

\node[box,below=of g2] (h2)
{PCA visualization + dendrogram};

\node[box,below=13mm of h2] (i)
{Cluster--class comparison};

\draw[arrow] (a)--(b);
\draw[arrow] (b)--(c);
\draw[arrow] (c)--(d);

\draw[arrow] (d)--(e1);
\draw[arrow] (d)--(e2);
\draw[arrow] (e2)--(f2);
\draw[arrow] (f2)--(g2);
\draw[arrow] (g2)--(h2);

\draw[arrow] (e1) |- (i);
\draw[arrow] (h2)--(i);

\end{tikzpicture}
}
\caption{Workflow of the statistical hardness analysis. Individual \textit{Swift}/XRT observations are classified using their hardness ratios and conservative lower and upper hardness bounds and are then used to derive $f_{\rm soft}$, $f_{\rm hard}$, and $CI$. These descriptors are used in two complementary analyses: a rule-based classification and unsupervised hierarchical clustering. The clustering uses only these three descriptors as input; the rule-based labels are introduced only after clustering to compare the resulting groups. Here, $HR$ denotes the hardness ratio, $S$ and $H$ are the soft- and hard-band count rates, respectively, $HR_{\rm lo}$ and $HR_{\rm hi}$ denote the conservative lower and upper hardness bounds, and $CI$ denotes the Crossing Index. PCA denotes principal component analysis and is used only for visualization, while CH and DB denote the Calinski--Harabasz and Davies--Bouldin indices, respectively.}
\label{fig:workflow}
\end{figure*}
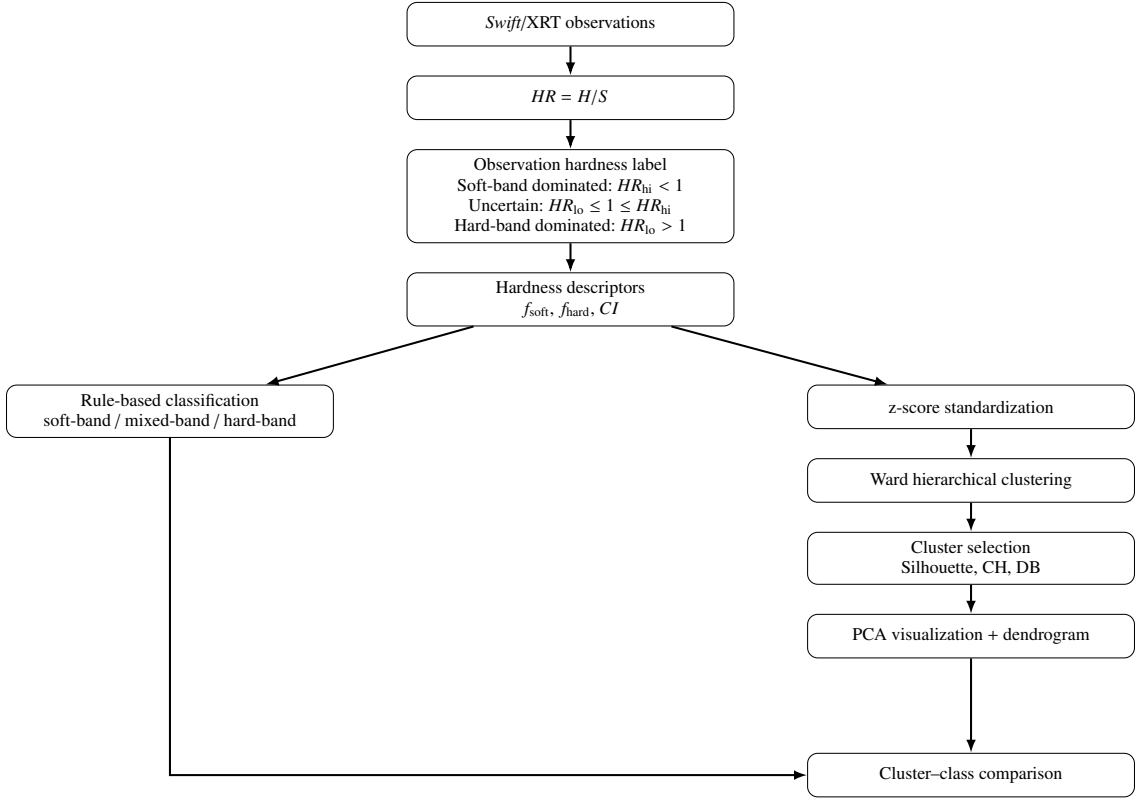

No dimensionality reduction was applied during the clustering procedure itself; all clustering analyses were performed in the original standardized three-dimensional parameter space. Principal Component Analysis (PCA) was used exclusively for visualization by projecting the standardized statistical parameter space onto two dimensions. In addition, a hierarchical dendrogram was constructed to illustrate the relationships among the sources and the hierarchical structure of the resulting clusters. The PCA projection and the corresponding dendrogram are presented in Figs.~\ref{fig:pca_clusters} and \ref{fig:dendrogram}, respectively. A schematic overview of the complete analysis workflow is shown in Fig.~\ref{fig:workflow}.

\subsection{Morphological analysis of hardness--intensity and Hard--Soft diagrams}

The hardness--intensity diagrams (HIDs) were examined to investigate whether the long-term spectral evolution of the 29 ULXs exhibits recurrent phenomenological patterns in the hardness--intensity plane. Because many sources display complex or non-linear behavior, a simple phenomenological approach was adopted rather than quantitative model fitting. The HIDs were therefore classified according to the overall morphology of their distributions in the hardness--intensity plane.

The Hard--Soft diagrams were analyzed to investigate the long-term relationship between the soft- and hard-band X-ray emission of each ULX. Although visual inspection provides a useful qualitative description of the observed morphologies, the strength of the coupling between the two energy bands requires quantitative statistical measurements. Therefore, for each source we calculated the Spearman rank correlation coefficient ($r_{\rm s}$) and its corresponding significance level ($p$-value) to quantify the monotonic correlation between the soft- and hard-band count rates. For each source, we fitted $\log_{10}H=a+\beta\log_{10}S$, where $S$ and $H$ are the soft- and hard-band count rates, using orthogonal distance regression with uncertainties in both coordinates. We included only observations with finite, positive count rates and count-rate uncertainties. The uncertainties in logarithmic space were calculated as $\sigma_{\log S}=\sigma_S/(S\ln 10)$ and $\sigma_{\log H}=\sigma_H/(H\ln 10)$. The reported slope uncertainty, $\sigma_{\beta}$, is the value returned by the regression for $\beta$. The resulting statistical parameters are listed in Table~\ref{tab:hs_statistics}.

\begin{table}
\caption{Spearman rank correlation coefficients and parameters of the log--log Hard--Soft fits.}
\label{tab:hs_statistics}
\centering
\begin{tabular}{lcccc}
\hline
Source & $r_{\rm s}$ & $p$-value & $\beta$ & $\sigma_{\beta}$ \\
\hline
Swift~J0243.6+6124 & 0.99 & $9.44\times10^{-57}$ & 1.08 & 0.01 \\
NGC~7793 P13 & 0.89 & $1.35\times10^{-143}$ & 0.94 & 0.02 \\
NGC~300 ULX--1 & 0.83 & $6.78\times10^{-18}$ & 0.94 & 0.07 \\
IC~10 X-1 & 0.82 & $1.13\times10^{-49}$ & 0.97 & 0.05 \\
SS433 & 0.79 & $1.14\times10^{-6}$ & 0.63 & 0.05 \\
M83 ULX--1 & 0.78 & $6.09\times10^{-14}$ & 1.13 & 0.11 \\
NGC~1313 X--2 & 0.76 & $2.30\times10^{-96}$ & 0.53 & 0.02 \\
NGC~1566 ULX-3 & 0.75 & $9.88\times10^{-7}$ & 1.02 & 0.09 \\
NGC~4559 X-10 & 0.65 & $1.74\times10^{-16}$ & 0.92 & 0.08 \\
NGC~300 X-1 & 0.64 & $1.05\times10^{-21}$ & 0.95 & 0.07 \\
NGC~4559 X-7 & 0.63 & $6.61\times10^{-22}$ & 0.93 & 0.05 \\
NGC~4258 X-3 & 0.62 & $1.02\times10^{-3}$ & 0.79 & 0.16 \\
NGC~3628 ULX--1 & 0.55 & $9.11\times10^{-4}$ & 0.64 & 0.08 \\
NGC~5907 ULX--1 & 0.53 & $1.05\times10^{-21}$ & 0.76 & 0.06 \\
M51 ULX-7 & 0.52 & $4.17\times10^{-16}$ & 0.71 & 0.06 \\
M51 ULX-8 & 0.51 & $3.40\times10^{-18}$ & 0.61 & 0.04 \\
NGC~247 ULX--1 & 0.50 & $9.24\times10^{-4}$ & 1.03 & 0.14 \\
NGC~3627 ULX--2 & 0.50 & $1.20\times10^{-3}$ & 0.82 & 0.11 \\
NGC~7456 ULX--1 & 0.48 & $1.91\times10^{-3}$ & 1.06 & 0.21 \\
NGC~5643 X-1 & 0.43 & $3.15\times10^{-5}$ & 0.84 & 0.10 \\
NGC~925 ULX-3 & 0.42 & $1.87\times10^{-3}$ & 1.49 & 0.25 \\
M33 X-7 & 0.41 & $9.13\times10^{-2}$ & 1.01 & 0.22 \\
NGC~1097 ULX-3 & 0.35 & $7.92\times10^{-2}$ & 0.64 & 0.15 \\
NGC~925 ULX--2 & 0.29 & $2.35\times10^{-4}$ & 0.29 & 0.05 \\
NGC~3521 ULX--1 & 0.28 & $2.16\times10^{-1}$ & 0.82 & 0.28 \\
NGC~4490 X8 & 0.25 & $1.40\times10^{-1}$ & 1.94 & 0.45 \\
M51 ULX--2 & 0.25 & $5.58\times10^{-2}$ & 1.07 & 0.24 \\
M51 ULX--eclipsing & 0.20 & $1.35\times10^{-1}$ & 1.64 & 0.35 \\
NGC~3623 ULX--1 & -0.30 & $4.33\times10^{-1}$ & -0.21 & 0.21 \\
\hline
\end{tabular}
\end{table}

\section{Results}

\subsection{Long-term X-ray variability}

Long-term \textit{Swift}/XRT light curves were constructed for all 29 ULXs to investigate their long-term X-ray variability. All sources exhibit substantial long-term flux variability over the \textit{Swift} monitoring baseline, although both the variability amplitude and temporal behavior differ markedly from source to source. Such long-term variability is a common characteristic of ULXs and is generally attributed to changes in the accretion flow, super-Eddington accretion, geometric effects, and, in some systems, orbital or super-orbital modulation. Consistent with previous studies, several sources exhibit the previously reported long-term or super-orbital variability. However, neither the Lomb--Scargle periodograms nor the bimodality analysis revealed any statistically significant new long-term periodicities or additional bimodal variability beyond those already reported in the literature.

\subsection{Statistical classification of the long-term hardness evolution}

The long-term hardness evolution of the 29 sources was quantified using the statistical framework described in Sect.~\ref{sec:hardness_classification}. For each source, the fractions of statistically soft-, hard-, and uncertain observations together with the CI were derived from the \textit{Swift}/XRT hardness-ratio measurements, allowing an objective classification into soft-band-dominated, hard-band-dominated, or mixed-band systems. The resulting statistical properties are summarized in Table~\ref{tab:hardness_classification}. As a complementary assessment of the structure formed by the same hardness descriptors, we performed an unsupervised hierarchical clustering analysis.

According to the rule-based classification, eight sources are classified as soft-band dominated, five as hard-band dominated, and the remaining 16 as mixed-band systems. The soft-band-dominated sources have no observed crossings between significant hardness labels ($CI=0$). NGC~247 ULX--1 has $f_{\rm soft}=100\%$, while M51 ULX--2, NGC~300 X--1, NGC~4559 X--7, and NGC~7456 ULX--1 have $f_{\rm hard}=0\%$. IC~10 X--1, NGC~3627 ULX--2, NGC~4490 X--8, SS433, and Swift~J0243.6+6124 are hard-band dominated, with $f_{\rm soft}=0\%$. The remaining 16 sources belong to the mixed-band class because they contain significant observations on both sides of the $HR=1$ boundary. Their CI values span a wide range, from $CI=2\%$ for NGC~5907 ULX-1 to $CI=56\%$ for NGC~925 ULX-3, demonstrating substantial diversity in the fraction of adjacent significant observations assigned to different hardness labels. The largest CI values are obtained for NGC~925 ULX-3 ($56\%$), NGC~4258 X-3 ($50\%$), NGC~5643 X-1 ($40\%$), NGC~925 ULX-2 ($32\%$), NGC~1097 ULX-3 ($31\%$), and M51 ULX-8 ($31\%$). These sources are assigned to Cluster~1, which has the largest mean CI among the three observational groups.

To investigate whether these three observational classes correspond to natural groupings within the statistical hardness parameter space, we performed an unsupervised hierarchical clustering analysis using only \(f_{\rm soft}\), \(f_{\rm hard}\), and \(CI\) as input features. The rule-based class labels were deliberately excluded from the clustering procedure and were used only after the clusters had been identified for comparison. Among the tested clustering solutions (\(k=2\)--6), the highest silhouette coefficient was obtained for the three-cluster solution (\(S=0.52\)). The corresponding Calinski--Harabasz and Davies--Bouldin indices were 39.77 and 0.664, respectively, further supporting the choice of a three-cluster solution. The first two principal components account for 94.5\% of the total variance, indicating that the original three-dimensional statistical hardness parameter space is well represented by the two-dimensional PCA projection shown in Fig.~\ref{fig:pca_clusters}. The hierarchical dendrogram shown in Fig.~\ref{fig:dendrogram} likewise supports the presence of three principal clusters.

The three clusters exhibit distinct statistical hardness properties (see Table \ref {tab:cluster_properties}). Cluster~1 is characterized by intermediate soft- and hard-band fractions and the largest mean CI ($\langle f_{\rm soft}\rangle=27.20\%$, $\langle f_{\rm hard}\rangle=11.60\%$, and $\langle CI\rangle=31.55\%$), and therefore represents the mixed-band group. Cluster~2 is dominated by hard-band observations ($\langle f_{\rm hard}\rangle=63.87\%$), with only a small soft-band contribution ($\langle f_{\rm soft}\rangle=2.86\%$) and a relatively low mean CI ($\langle CI\rangle=6.20\%$), representing the predominantly hard-band group. Cluster~3 is dominated by soft-band observations ($\langle f_{\rm soft}\rangle=72.56\%$), with a negligible hard-band fraction ($\langle f_{\rm hard}\rangle=0.22\%$) and a mean CI of $0.56\%$, representing the predominantly soft-band group.

\begin{table}
\caption{Mean statistical hardness properties of the three clusters identified by hierarchical clustering.}
\label{tab:cluster_properties}
\centering
\begin{tabular}{ccccc}
\hline\hline
Cluster & Soft (\%) & Hard (\%) & CI (\%) & Sources \\
\hline
1 & 27.20 & 11.60 & 31.55 & 10 \\
2 & 2.86 & 63.87 & 6.20 & 10 \\
3 & 72.56 & 0.22 & 0.56 & 9 \\
\hline
\end{tabular}
\end{table}

Comparison with the rule-based statistical classification (Table~\ref{tab:cluster_comparison}) demonstrates a high degree of consistency between the two complementary approaches. All ten sources assigned to Cluster~1 are classified as mixed-band by the rule-based statistical classification. Cluster~2 is dominated by hard-band dominated systems but also includes five sources that are classified as mixed-band by the rule-based scheme (NGC~300 ULX--1, NGC~1313 X--2, NGC~3628 ULX--1, NGC~5907 ULX--1, and NGC~7793 P13), whereas Cluster~3 is dominated by soft-band dominated systems and contains only one source classified as mixed-band by the rule-based classification (M83 ULX--1). Overall, only six sources are assigned to different groups by the two methods. These discrepancies arise because the rule-based classification applies fixed decision thresholds, whereas the unsupervised hierarchical clustering considers the combined distribution of all statistical hardness parameters. The agreement between the two approaches shows that the adopted descriptors form a coherent three-group structure within the observed-count parameter space. However, because both approaches are based on the same hardness descriptors, this agreement does not independently demonstrate that the groups represent physically distinct ULX populations.

\begin{table}
\caption{Comparison between the rule-based statistical classification and the hierarchical clustering results.}
\label{tab:cluster_comparison}
\centering
\begin{tabular}{lccc}
\hline\hline
 & Soft-band & Hard-band & Mixed-band \\
\hline
Cluster 1 & 0 & 0 & 10 \\
Cluster 2 & 0 & 5 & 5 \\
Cluster 3 & 8 & 0 & 1 \\
\hline
\end{tabular}
\tablefoot{Six sources are assigned to different groups by the two approaches, indicating a high degree of consistency between the rule-based statistical classification and the unsupervised hierarchical clustering analysis.}
\end{table}

\subsection{Statistical properties of the Hard--Soft relation}

To quantify the coupling between the soft- and hard-band emission, the Spearman rank correlation coefficient ($r_{\rm s}$), its corresponding significance level ($p$-value), and the linear slope ($\beta$) were calculated for each source. The resulting statistical parameters are presented in Table~\ref{tab:hs_statistics}. With the exception of NGC~3623 ULX--1, all sources exhibit positive correlations between the soft- and hard-band count rates, with Spearman coefficients ranging from $-0.30$ to $0.99$. The strongest correlations are measured for Swift~J0243.6+6124 ($r_{\rm s}=0.99$), NGC~300 ULX--1 ($r_{\rm s}=0.83$), IC~10 X--1 ($r_{\rm s}=0.82$), SS433 ($r_{\rm s}=0.79$), M83 ULX--1 ($r_{\rm s}=0.78$), and NGC~1313 X--2 ($r_{\rm s}=0.76$), indicating a strong long-term coupling between the soft- and hard-band emission. In contrast, M33 X--7, M51 ULX--2, M51 ULX--eclipsing, NGC~1097 ULX--3, NGC~3521 ULX--1, and NGC~4490 X8 exhibit statistically insignificant correlations ($p>0.05$). The fitted slopes span a wide range, from $\beta=-0.21$ for NGC~3623 ULX--1 to $\beta=1.94$ for NGC~4490 X8. Most sources have slopes close to unity, indicating comparable long-term variations in the soft- and hard-band count rates, whereas flatter ($\beta<1$) or steeper ($\beta>1$) relations suggest different relative responses of the two energy bands during the long-term spectral evolution.

\section{Discussion}

The rule-based classification and hierarchical clustering assign 23 of the 29 sources to corresponding observational groups, while six sources receive different assignments. All discrepancies occur between a persistent group and the mixed-band group; no source is reassigned directly between the hard-band-dominated and soft-band-dominated groups. This pattern reflects the different criteria used by the two approaches. The rule-based method assigns a source to the mixed-band class whenever significant observations occur on both sides of $HR=1$, whereas the clustering considers the joint distribution of $f_{\rm soft}$, $f_{\rm hard}$, and CI. Consequently, M83 ULX-1 is associated with the soft-band-dominated cluster, while NGC~300 ULX-1, NGC~1313 X-2, NGC~3628 ULX-1, NGC~5907 ULX-1, and NGC~7793 P13 are associated with the hard-band-dominated cluster despite being classified as mixed-band by the rule-based method. These discrepancies show that the two approaches emphasize different aspects of the same observed-count descriptor space.

The hierarchical clustering provides a complementary characterization of the structure formed by $f_{\rm soft}$, $f_{\rm hard}$, and CI. Because these descriptors also underlie the rule-based classification, the clustering should not be regarded as an independent physical validation. The agreement between the two methods instead shows that the adopted descriptors form a coherent statistical structure in the observed-count parameter space. The six discrepant assignments indicate that the boundaries between the observational groups are gradual, and that the classification of sources near these boundaries depends on whether the presence of any significant crossing or the overall distribution of the three descriptors is emphasized.

We compared the locations of confirmed and candidate pulsating ULXs, black-hole systems, and sources with unidentified compact objects in the descriptor space only after completing the statistical classification. Some confirmed and candidate pulsating ULXs occur in the predominantly hard-band group, while several systems with unidentified compact objects occupy the mixed-band group. However, the sample is small, the compact-object identifications are heterogeneous, and several sources lie near the cluster boundaries. This post hoc comparison therefore does not establish a unique association between the observed \textit{Swift}/XRT groups and compact-object type. At most, it identifies empirical trends that can be examined using larger samples and independent physical measurements.

A clear empirical pattern emerges from the rule-based classification: all four sources in our sample with reported long-term X-ray periodic modulation discussed in an orbital or super-orbital context---NGC~5907 ULX-1, M51 ULX-7, NGC~925 ULX-3, and NGC~7793 P13---are classified as mixed-band systems. Thus, every periodically modulated source in the sample exhibits significant \textit{Swift}/XRT observations on both sides of the $HR=1$ boundary. This common classification is consistent with a scenario in which long-term geometric modulation, potentially associated with precession of the supercritical accretion flow or wind, changes the relative contributions of the soft- and hard-band emission. The presence of all four periodically modulated sources in the rule-based mixed-band class is an observational result that may help prioritize targets for follow-up searches for long-term modulation. However, the small number of such sources and the uncertain physical origin of some reported periods preclude using mixed-band hardness behavior as a reliable indicator of super-orbital modulation.

The physical interpretation of the reported periods must nevertheless be considered on a source-by-source basis. NGC~5907 ULX-1 exhibits an approximately 78~d X-ray modulation interpreted as super-orbital \citep{2016ApJ...827L..13W}, while M51 ULX-7 shows an approximately 38--39~d super-orbital modulation \citep{2020MNRAS.491.4949V,2022ApJ...925...18B}. The approximately 126~d modulation of NGC~925 ULX-3 may have either an orbital or super-orbital origin \citep{2022MNRAS.512.1814S}. NGC~7793 P13 shows more complex long-term behavior involving its approximately 65~d X-ray and optical modulations, orbital solution, and long-term phase evolution \citep{2017ApJ...835L...9H,2018A&A...616A.186F,2021A&A...651A..75F}. Accordingly, the concentration of all four periodically modulated sources in the rule-based mixed-band class is robust as an observational result, whereas its specific interpretation as an association with super-orbital variability depends on the physical origin of the individual periods.

The four mixed-band sources also display different degrees of hardness-label alternation. NGC~5907 ULX-1 and NGC~7793 P13 have CI values of only approximately 0.02 and 0.04, respectively, showing that their mixed-band classifications result from a small number of crossings over the full monitoring baseline. Their high hard-band fractions also place them in the predominantly hard-band group in the hierarchical clustering. This does not alter their rule-based mixed-band classifications; instead, it shows that the mixed-band class includes both sources with frequent hardness-label alternation and predominantly hard-band sources that cross $HR=1$ only occasionally. We therefore retain the concentration of the four periodically modulated sources in the rule-based mixed-band class as an important empirical result, while treating a direct physical relation between super-orbital variability and mixed-band behavior as suggestive rather than established.

While the Spearman coefficient ($r_{\rm s}$) measures the degree of coupling between the hard- and soft-band emission, the linear slope ($\beta$) describes their relative long-term variability. Values of $\beta \approx 1$ indicate that the hard- and soft-band emission vary proportionally, suggesting that both spectral components respond similarly to changes in the accretion flow. By contrast, $\beta < 1$ implies that the soft-band emission exhibits larger relative variations than the hard component, whereas $\beta > 1$ indicates a stronger response of the hard-band emission. Consequently, sources with similar values of $r_{\rm s}$ may still exhibit substantially different long-term spectral evolution because the relative amplitudes of the hard and soft variability are different. The widest distribution of $\beta$ is observed among the mixed-band group. This indicates that, although these sources contain significant observations on both sides of the $HR=1$ boundary, they do not follow a common mode of spectral evolution. Instead, some mixed-band ULXs are dominated by larger soft-band variations, whereas others show stronger hard-band variability. This diversity is consistent with the mixed-band group representing a heterogeneous mixture of accretion geometries, viewing angles, and wind properties rather than a single physical class. 

Interestingly, the dynamically confirmed black-hole systems occupy a narrow range around $\beta \approx 1$, indicating nearly proportional long-term evolution of the hard- and soft-band emission. In contrast, neutron-star ULXs and, particularly, ULXs with unidentified compact objects span a much broader range of $\beta$. Although the present sample is too small to establish statistically significant differences between compact-object populations, this trend suggests that proportional Hard--Soft evolution may be more common among the black-hole systems included in the present sample. The combination of $r_{\rm s}$ and $\beta$ provides considerably more information than either parameter alone. Two ULXs may exhibit similarly strong Hard--Soft correlations while displaying very different values of $\beta$, indicating that the hard and soft emission evolve coherently but with different relative amplitudes. Thus, $r_{\rm s}$ quantifies the strength of the coupling between the two energy bands, whereas $\beta$ characterizes how that coupling is realized. Together with the statistical hardness fractions and the Crossing Index, these parameters provide a more complete statistical description of the long-term spectral evolution of ULXs.

The hardness--intensity diagrams indicate that ULXs do not follow a common track in the hardness--intensity plane. Unlike Galactic black-hole X-ray binaries, none of the candidate black-hole systems in the present sample exhibits a well-defined hysteretic q-shaped track. This behavior is consistent with the current picture of ULXs as super-Eddington accretors. The statistical hardness populations identified in this work also exhibit distinct hardness--intensity morphologies. Persistent hard-band- and soft-band-dominated sources generally exhibit more coherent hardness--intensity evolution than mixed-band systems. In contrast, mixed-band ULXs frequently occupy broad regions of the hardness--intensity plane, with substantially different hardness ratios observed at comparable X-ray intensities. This behavior indicates that the spectral hardness is not determined by the X-ray intensity alone. Additional physical processes, such as changes in the accretion geometry or super-Eddington outflows, are therefore likely to influence the observed spectral evolution. The HID morphologies provide complementary descriptive information, showing that sources assigned to the mixed-band group generally occupy broader regions of the observed hardness--intensity plane. This comparison does not constitute an independent validation of physically distinct source populations.

To assess the dependence of the hierarchical clustering on the limited sample size, we performed leave-one-out and bootstrap stability tests using the same standardized parameter space defined by $f_{\rm soft}$, $f_{\rm hard}$, and CI. In each leave-one-out trial, one source was removed, the remaining 28 sources were standardized again, and solutions with $k=2$--6 were reevaluated using the silhouette coefficient. The three-cluster solution remained preferred in 25 of 29 trials (86.2\%). The four exceptions occurred when M83 ULX-1, NGC 5643 X-1, NGC 5907 ULX-1, or NGC 7793 P13 was omitted; for the latter two trials, however, the differences between the $k=3$ and $k=4$ silhouette coefficients were only 0.0035 and 0.0001, respectively. This sensitivity reflects their locations in the parameter space: M83 ULX-1 and NGC 5643 X-1 influence the soft- and transitional-group geometries, respectively, whereas the confirmed pulsating ULXs NGC 5907 ULX-1 and NGC 7793 P13 form a closely spaced pair in the hard-band dominated group. We additionally generated 10,000 bootstrap samples by resampling the 29 sources with replacement and reclustered each sample with $k=3$. The mean cluster-wise Jaccard similarities were 0.868 for the hard-band dominated group, 0.868 for the soft-band dominated group, and 0.782 for the transitional group. Thus, the persistent hard- and soft-band dominated groups are highly stable against changes in sample composition, whereas the transitional group is comparatively more sensitive. The lowest bootstrap membership-retention fractions were obtained for NGC 1313 X-2, M33 X-7, NGC 300 ULX-1, and M51 ULX-eclipsing, with values of 42.0\%, 49.4\%, 68.8\%, and 70.5\%, respectively. Most of the remaining sources retained their original memberships in more than 90\% of the bootstrap samples. These results support the statistical separation of the three observational groups while identifying individual sources whose assignments depend more strongly on the composition of the sample; such assignments should therefore not be interpreted as evidence for sharply distinct physical populations.

\subsection{Effect of absorption and continuum shape on the hardness classification}
\label{sec:absorption_test}

\begin{figure}
\resizebox{\hsize}{!}{\includegraphics{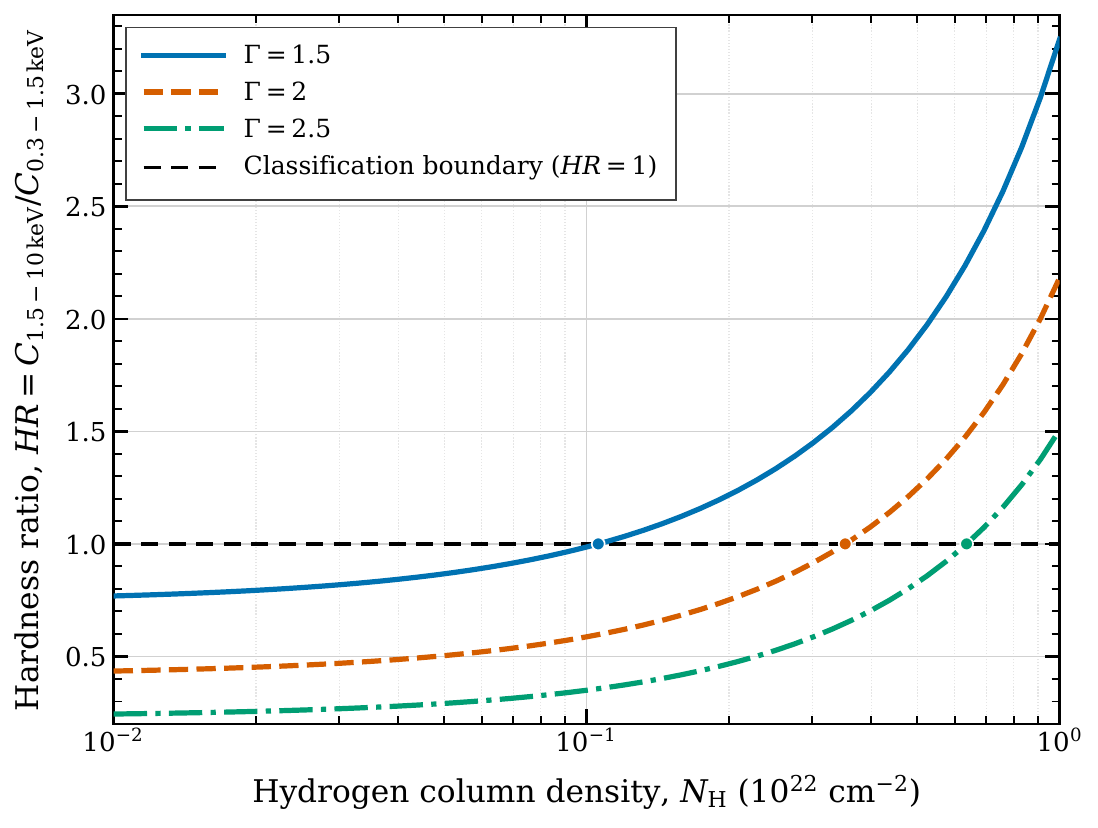}}
\caption{Response-folded \textit{Swift}/XRT hardness ratio as a function of absorbing column for absorbed power-law models with photon indices $\Gamma=1.5$, 2.0, and 2.5. Model count rates were calculated in the 0.3--1.5 and 1.5--10~keV bands using a representative \textit{Swift}/XRT PC-mode response. The horizontal dashed line marks the adopted $HR=1$ boundary, while the colored circles indicate the absorbing column at which each model crosses this boundary.}
\label{fig:nh_hr}
\end{figure}

Photoelectric absorption preferentially suppresses soft X-rays, while the intrinsic continuum slope determines the relative number of photons detected in the soft and hard bands. Both effects may therefore influence the observed hardness classification. We quantified their contributions by forward-folding \texttt{TBabs*powerlaw} models through the \textit{Swift}/XRT PC-mode response of NGC~1313~X-2 (ObsID 00036555002), without fitting the observed spectrum. We adopted the abundances of \citet{2000ApJ...542..914W} and the photoelectric cross sections of \citet{1996ApJ...465..487V}, and considered photon indices of $\Gamma=1.5$, 2.0, and 2.5. For each photon index, $N_{\rm H}$ was varied over 51 logarithmically spaced values between $0.01$ and $1.0\times10^{22}\ {\rm cm^{-2}}$. The power-law normalization was fixed at $10^{-4}\ {\rm photons\ keV^{-1}\ cm^{-2}\ s^{-1}}$ at 1~keV. Predicted count rates were calculated in the 0.3--1.5 and 1.5--10~keV bands, and the modeled hardness was determined as $HR_{\rm mod}=C_{\rm 1.5-10\,keV}/C_{\rm 0.3-1.5\,keV}$. As shown in Fig.~\ref{fig:nh_hr}, increasing $N_{\rm H}$ at fixed $\Gamma$ shifts the model toward higher hardness, whereas decreasing $\Gamma$ at fixed $N_{\rm H}$ produces a similar effect by flattening the continuum. The models cross $HR=1$ at $N_{\rm H}\simeq0.106$, 0.352, and $0.636\times10^{22}\ {\rm cm^{-2}}$ for $\Gamma=1.5$, 2.0, and 2.5, respectively. The position of the hardness boundary is therefore determined by the combined effects of absorption and intrinsic continuum shape.

Our classification was designed to compare the observed long-term hardness evolution of ULXs measured with the same detector, energy bands, and processing procedure. The measured count rates retain the combined effects of each source's continuum shape, line-of-sight absorption, and spectral variability. Accordingly, a hard-band-dominated observation may result from a flatter continuum, stronger absorption, or both, and these contributions cannot be separated using count-based hardness alone. The clustering analysis tests whether these observed \textit{Swift}/XRT properties form recurring patterns across the sample rather than providing absorption-corrected spectral classifications. The same \textit{Swift}/XRT energy bands and hardness definition have been used to identify distinct long-term regimes in NGC~1313~X-2 \citep{2024MNRAS.531.3118G}, while detailed \textit{XMM-Newton} spectroscopy showed that its different \textit{Swift} regimes can correspond to distinct intrinsic spectral shapes \citep{2022MNRAS.511.5346S}. The resulting hard-band-dominated, soft-band-dominated, and mixed groups should therefore be interpreted as similarities in long-term \textit{Swift}/XRT behavior, with their physical interpretation subject to the joint dependence of hardness on $N_{\rm H}$ and continuum shape.

\subsection{Sensitivity of the Crossing Index to sampling}
\label{sec:ci_sampling}

We tested the sensitivity of the Crossing Index to the number of available observations using four well-monitored sources that span different hardness patterns: NGC~300~X-1, NGC~925~ULX-2, NGC~925~ULX-3, and NGC~4559~X-10 (see Table \ref{tab:ci_downsampling}). For each source, we randomly retained between 20\% and 85\% of the original observations while preserving their chronological order. Uncertain observations were removed before calculating the CI, following the procedure used in the main analysis. At each retention fraction, the random selection and CI calculation were repeated $10^{5}$ times.

No dramatic or uniform systematic change in CI was found. When 50\% of the observations were retained, the median CI remained unchanged at 0 for NGC~300~X-1, decreased from 0.563 to 0.500 for NGC~925~ULX-3, increased from 0.320 to 0.417 for NGC~925~ULX-2, and increased from 0.250 to 0.269 for NGC~4559~X-10. Thus, the largest absolute change at this sampling level was approximately 0.10. Even when only 20\% of the observations were retained, the qualitative behavior of the sources was preserved: NGC~300~X-1 remained at $CI=0$, while the mixed sources retained nonzero median CI values. The main effect of reduced sampling was a broadening of the CI distributions rather than a substantial change in their central values. We therefore retain CI as a descriptive measure of the fraction of consecutive significant observations assigned to different hardness states, while avoiding its interpretation as a time-dependent transition frequency. CI values for sparsely sampled sources should be regarded as less precisely constrained.

\begin{table}
\centering
\caption{Sensitivity of the Crossing Index to random downsampling.}
\label{tab:ci_downsampling}
\begin{tabular}{lcccc}
\hline\hline
Source & Original & 75\% & 50\% & 20\% \\
\hline
NGC~300~X-1 & 0.00 & 0.00 & 0.00 & 0.00 \\
NGC~925~ULX-2 & 0.32 & 0.36 & 0.42 & 0.46 \\
NGC~925~ULX-3 & 0.56 & 0.54 & 0.50 & 0.50 \\
NGC~4559~X-10 & 0.25 & 0.26 & 0.27 & 0.28 \\
\hline
\end{tabular}
\tablefoot{Values are the median CI from $10^{5}$ random realizations at each retention fraction. Observations were randomly selected while retaining their chronological order, and uncertain observations were excluded as in the main analysis.}
\end{table}

\subsection{Robustness of the classification in the low-count regime}
\label{sec:bayesian_hr}

To test the effect of non-Gaussian uncertainties at low count levels, we repeated the hardness classification for NGC~925 ULX-3 using the 53 observations included in the original analysis and a Bayesian Poisson source--background model following the statistical framework of \citet{2006ApJ...652..610P}. These observations contain 7--47 events in the source aperture over 0.3--10~keV, with a median of 23 events. For each observation, the source-aperture counts $n_j$ and background-region counts $m_j$ were extracted independently in the soft ($j=S$, 0.3--1.5~keV) and hard ($j=H$, 1.5--10~keV) bands from the \textit{Swift}/XRT event lists. The background scaling factor was calculated as $\alpha_j=A_{{\rm src},j}/A_{{\rm bg},j}$ from the corresponding extraction areas. We adopted the Poisson source--background model $n_j\sim{\rm Poisson}(\lambda_j+\alpha_j\beta_j)$ and $m_j\sim{\rm Poisson}(\beta_j)$, where $\lambda_j$ is the expected net source intensity and $\beta_j$ is the expected number of background-region events. Independent uniform priors were assigned to the non-negative $\lambda_j$ and $\beta_j$ parameters. The resulting marginal posterior for each $\lambda_j$ was evaluated using its finite gamma-mixture representation, and $2\times10^{5}$ posterior draws were generated for each band and observation. The posterior hardness ratio was then calculated as $HR_{\rm post}=\lambda_H/\lambda_S$, thereby propagating the Poisson uncertainties in both the source and background measurements without applying a Gaussian approximation to their ratio. Following the one-sided probability corresponding to the original $1\sigma$ criterion, an observation was classified as hard-band dominated when $P(HR_{\rm post}>1)\geq0.8413$, as soft-band dominated when $P(HR_{\rm post}<1)\geq0.8413$, and as uncertain otherwise. We also repeated the classification using a probability threshold of 0.90 to test its dependence on this choice.

The Bayesian analysis changed the labels of 13 of the 53 observations. The numbers of soft-band-dominated, hard-band-dominated, and uncertain observations changed from 11, 6, and 36 to 4, 10, and 39, respectively, corresponding to changes in $(f_{\rm soft},f_{\rm hard},f_{\rm unc})$ from $(20.8\%,11.3\%,67.9\%)$ to $(7.5\%,18.9\%,73.6\%)$. The CI decreased from 0.563 to 0.385. Nevertheless, NGC~925 ULX-3 retained its mixed-band classification because significant observations remained on both sides of $HR=1$; the same source-level classification was obtained with the more conservative probability threshold of 0.90. Thus, the treatment of low-count uncertainties affects some individual labels and the numerical value of CI, but does not change the broader classification of this representative low-count source.

\section{Conclusions}

We presented a homogeneous statistical characterization of the long-term observed hardness behavior of 29 ULXs using more than 3500 archival \textit{Swift}/XRT observations. Individual observations were classified as soft-band dominated, hard-band dominated, or uncertain, and each source was described using $f_{\rm soft}$, $f_{\rm hard}$, and the newly introduced CI. CI measures the fraction of adjacent significant observations located on opposite sides of $HR=1$ and should not be interpreted as a transition rate or physical variability timescale.

The rule-based analysis classifies eight sources as soft-band dominated, five as hard-band dominated, and 16 as mixed-band systems. Complementary hierarchical clustering recovers the corresponding three-group structure in the observed-count descriptor space. The three-cluster solution remained preferred in 25 of 29 leave-one-out trials, while bootstrap tests showed that the persistent hard- and soft-band groups are more stable than the mixed-band group. Agreement between the two approaches demonstrates coherent observational structure in the adopted descriptors but does not establish that the groups constitute physically distinct ULX populations.

Response-folding calculations demonstrate that absorption can shift a source across the $HR=1$ boundary and that the column density required to do so depends strongly on the continuum slope. A Bayesian Poisson source--background reanalysis of a representative low-count source reduced its CI from 56.3\% to 38.5\% but preserved its mixed-band classification. Random downsampling of four well-monitored sources produced no common directional shift in CI; at 50\% retention, the largest absolute change in the median CI was approximately 0.10. These tests show that absorption, low-count statistics, and temporal sampling affect the numerical descriptors, while the source-level classifications examined in the robustness tests remained qualitatively stable.

Sources with known or candidate compact-object identifications preferentially occupy the persistent band-dominated groups, whereas systems with unidentified compact objects occur more frequently in the mixed-band group. In addition, all four sources whose reported long-term X-ray modulations have been discussed in an orbital or super-orbital context---NGC~5907 ULX-1, M51 ULX-7, NGC~925 ULX-3, and NGC~7793 P13---belong to the rule-based mixed-band class. NGC~5907 ULX-1 and NGC~7793 P13 nevertheless have low CI values and are assigned to the predominantly hard-band group by the clustering analysis, showing that mixed-band classification does not necessarily imply frequent hardness-label alternation. The concentration of all four periodically modulated sources in the rule-based mixed-band class is an important empirical result, although its specific physical connection with super-orbital variability remains to be confirmed. The proposed framework therefore provides a reproducible description of persistent and mixed-band long-term \textit{Swift}/XRT behavior. Whether mixed-band behavior can help prioritize searches for long-term modulation remains to be tested with larger samples.

\begin{acknowledgements}
This work is partially supported by the Bundesministerium f\"ur Wirtschaft und Energie through the Deutsches Zentrum f\"ur Luft- und Raumfahrt e.V. (DLR) under the grant 50 OR 2517. W.Y. acknowledges support from the Alexander von Humboldt Foundation. We thank the anonymous referee for their careful reading of the manuscript and constructive comments, which helped improve the clarity of this work.
\end{acknowledgements}

\bibpunct{(}{)}{;}{a}{}{,}
\bibliographystyle{statistical}
\bibliography{statistical}

\begin{appendix}
\section{Additional \textit{Swift}/XRT diagnostic diagrams} \label{figures in appen}

\begin{figure}
\resizebox{\hsize}{!}{\includegraphics{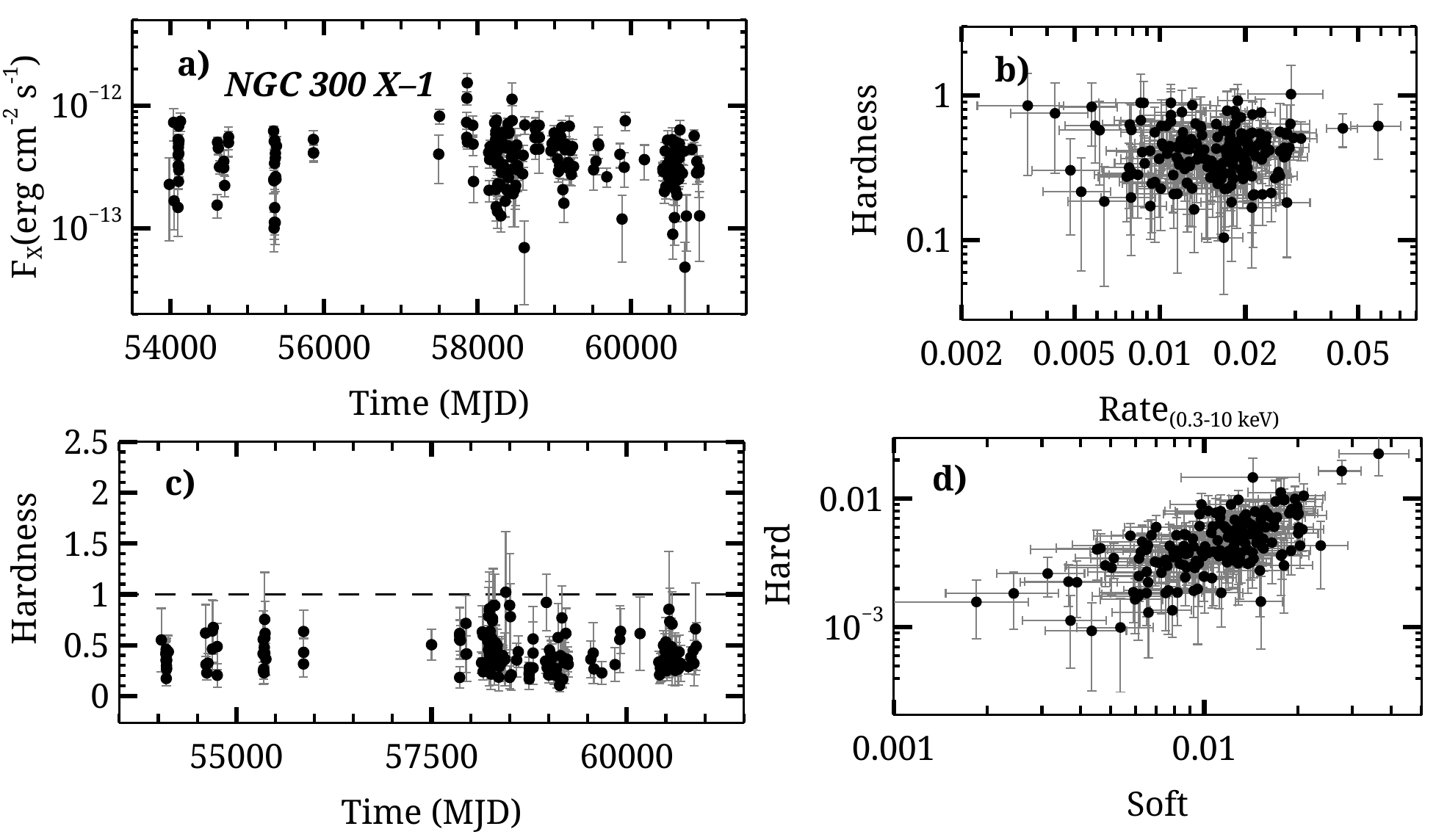}}
\caption{Same as Fig.~\ref{fig:swiftj0243}, but for NGC 300 X-1.}
\label{ngc300}
\end{figure}

\begin{figure}
\resizebox{\hsize}{!}{\includegraphics{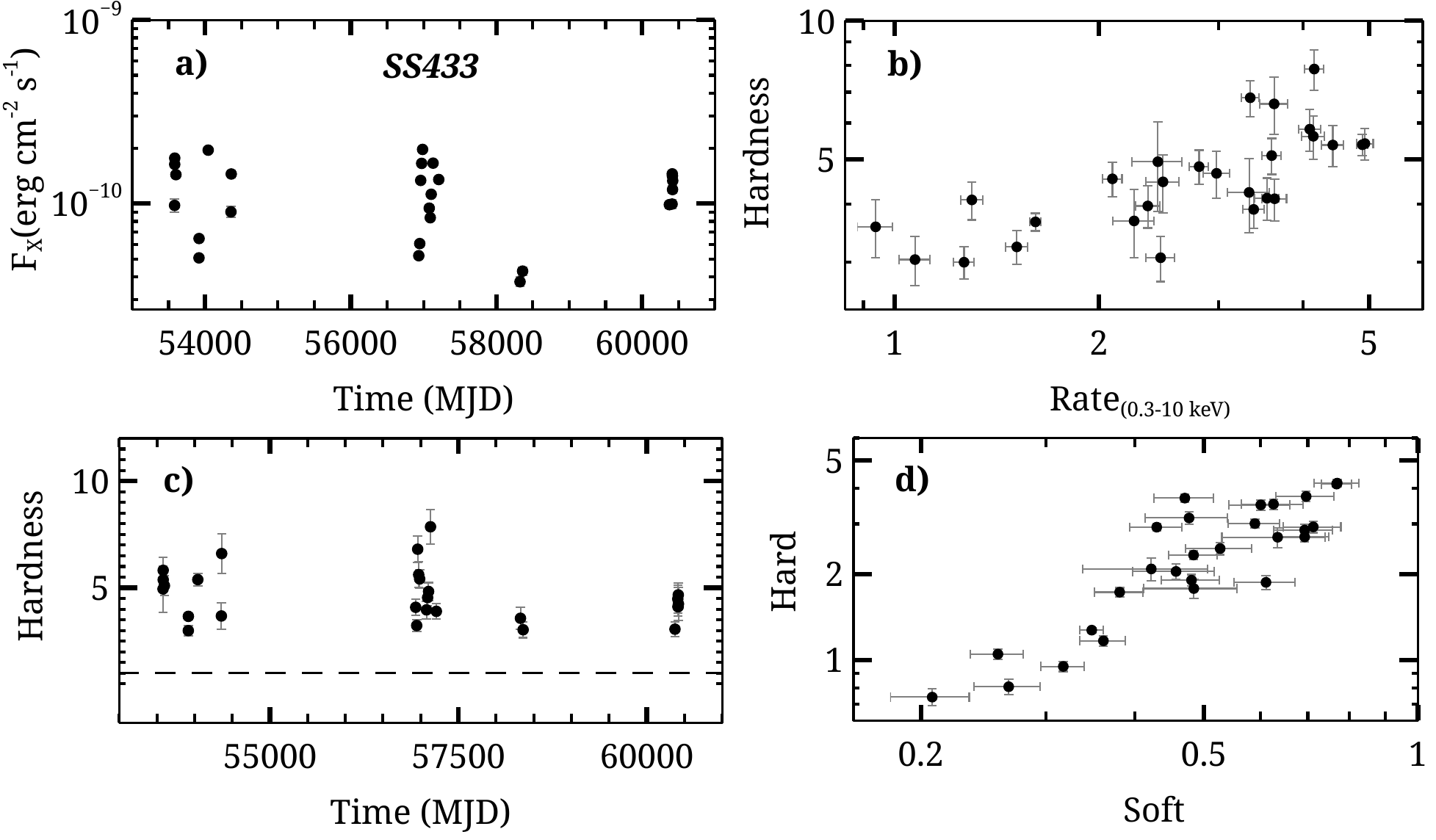}}
\caption{Same as Fig.~\ref{fig:swiftj0243}, but for SS433.}
\label{SS433}
\end{figure}

\begin{figure}
\resizebox{\hsize}{!}{\includegraphics{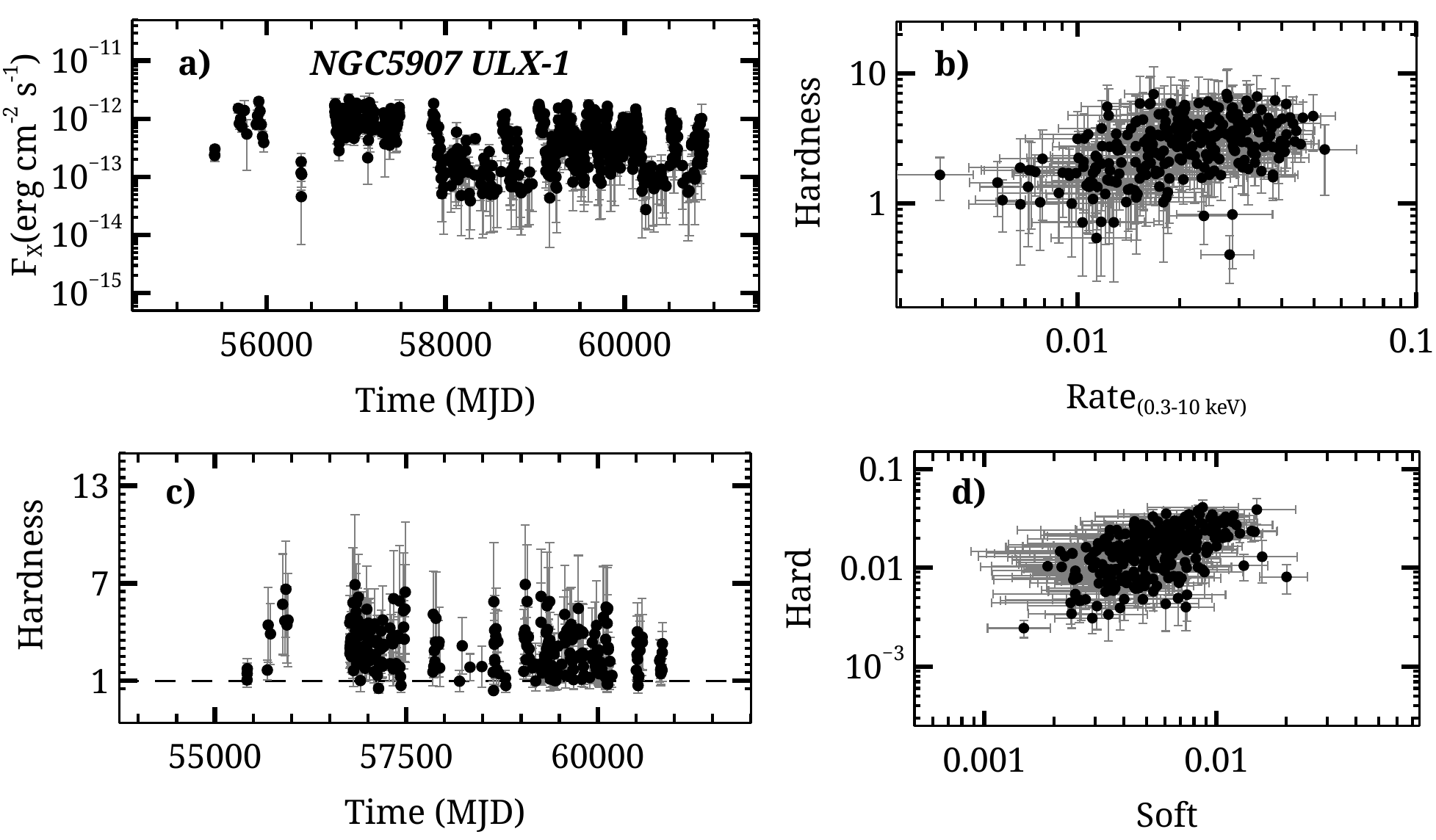}}
\caption{Same as Fig.~\ref{fig:swiftj0243}, but for NGC 5907 ULX--1.}
\label{NGC 5907}
\end{figure}

\begin{figure}
\resizebox{\hsize}{!}{\includegraphics{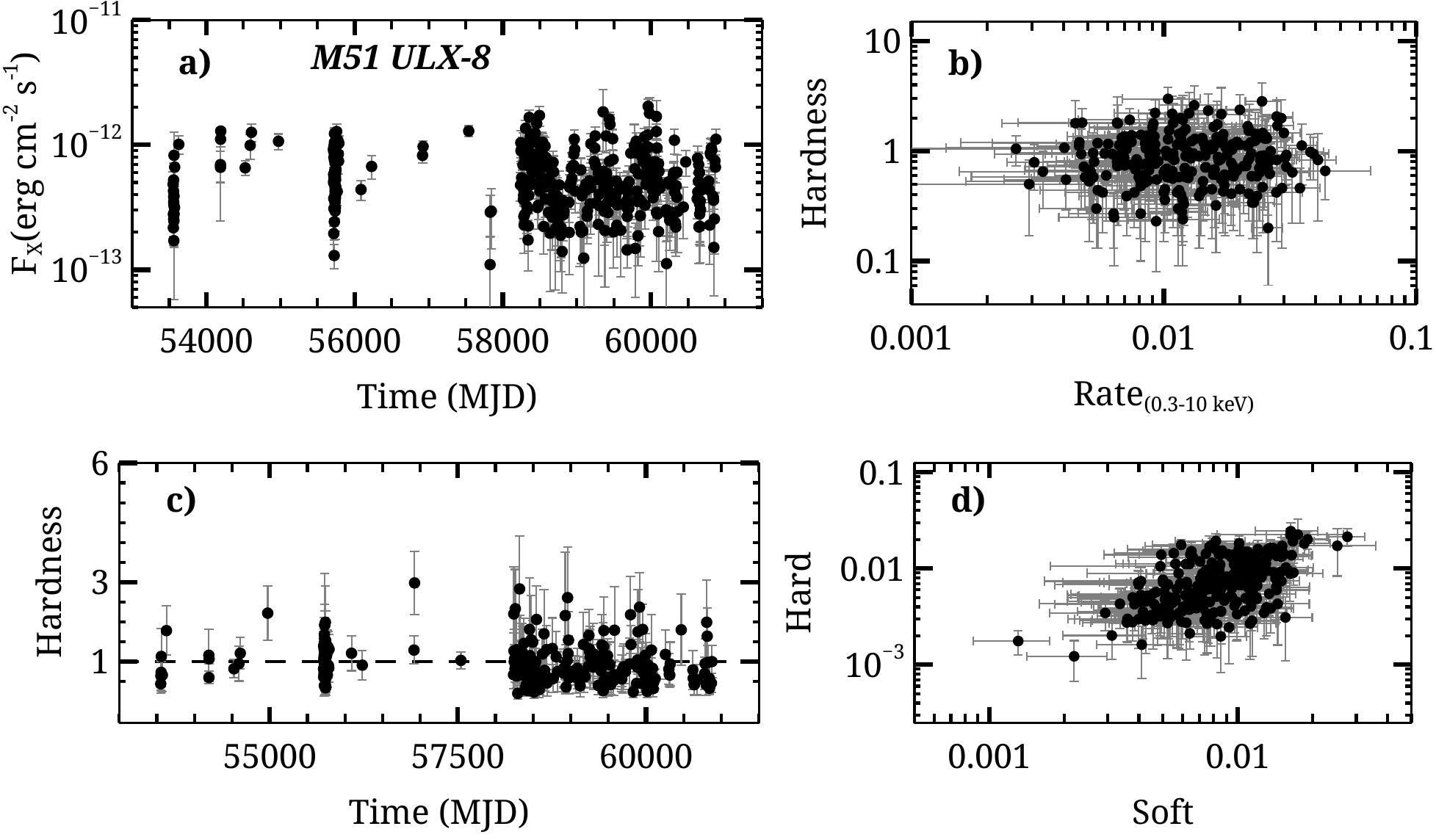}}
\caption{Same as Fig.~\ref{fig:swiftj0243}, but for M51 ULX-8.}
\label{M51 ULX-8}
\end{figure}

\begin{figure}
\resizebox{\hsize}{!}{\includegraphics{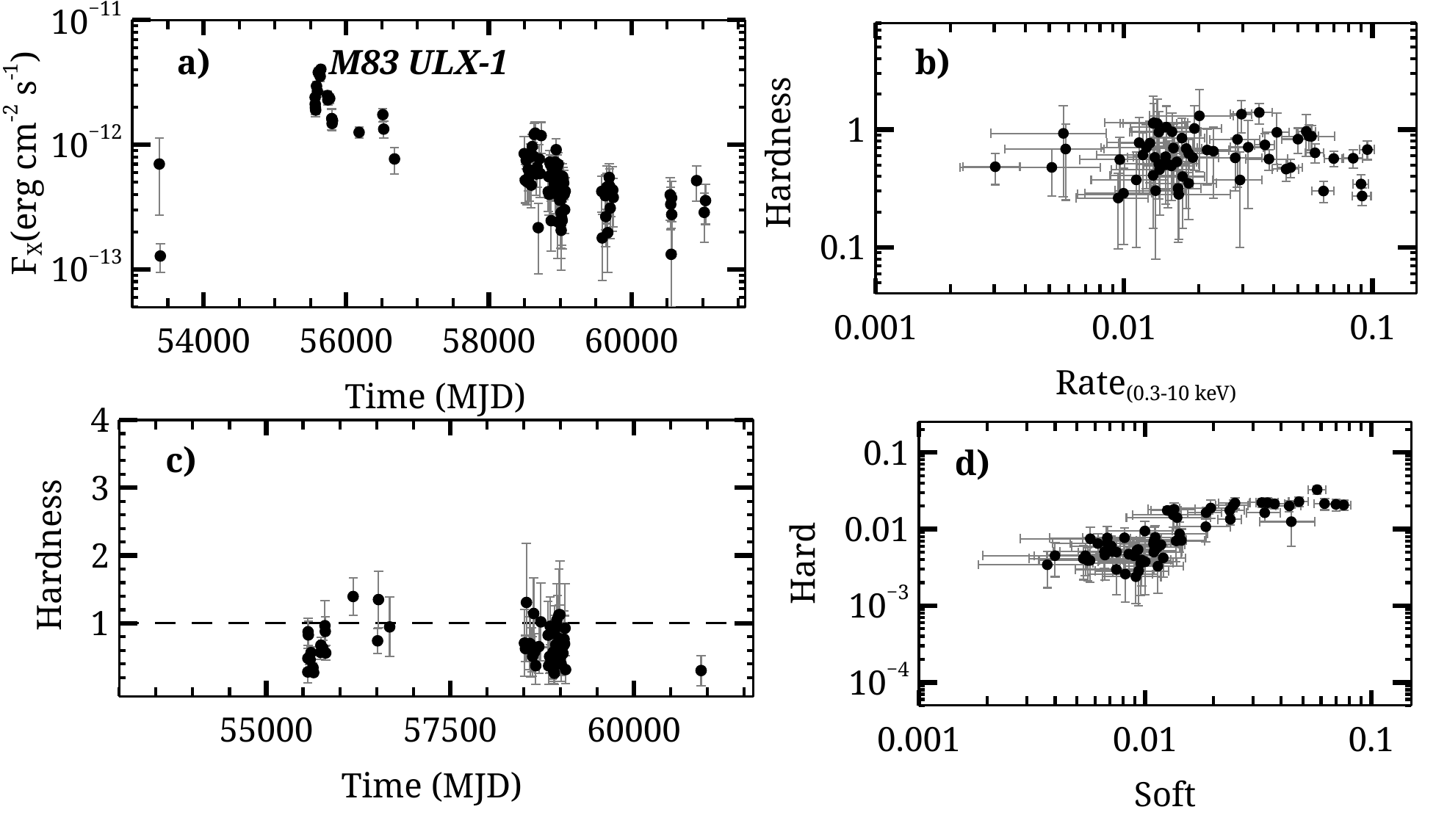}}
\caption{Same as Fig.~\ref{fig:swiftj0243}, but for M83 ULX--1.}
\label{M83ULX1}
\end{figure}

\begin{figure}
\resizebox{\hsize}{!}{\includegraphics{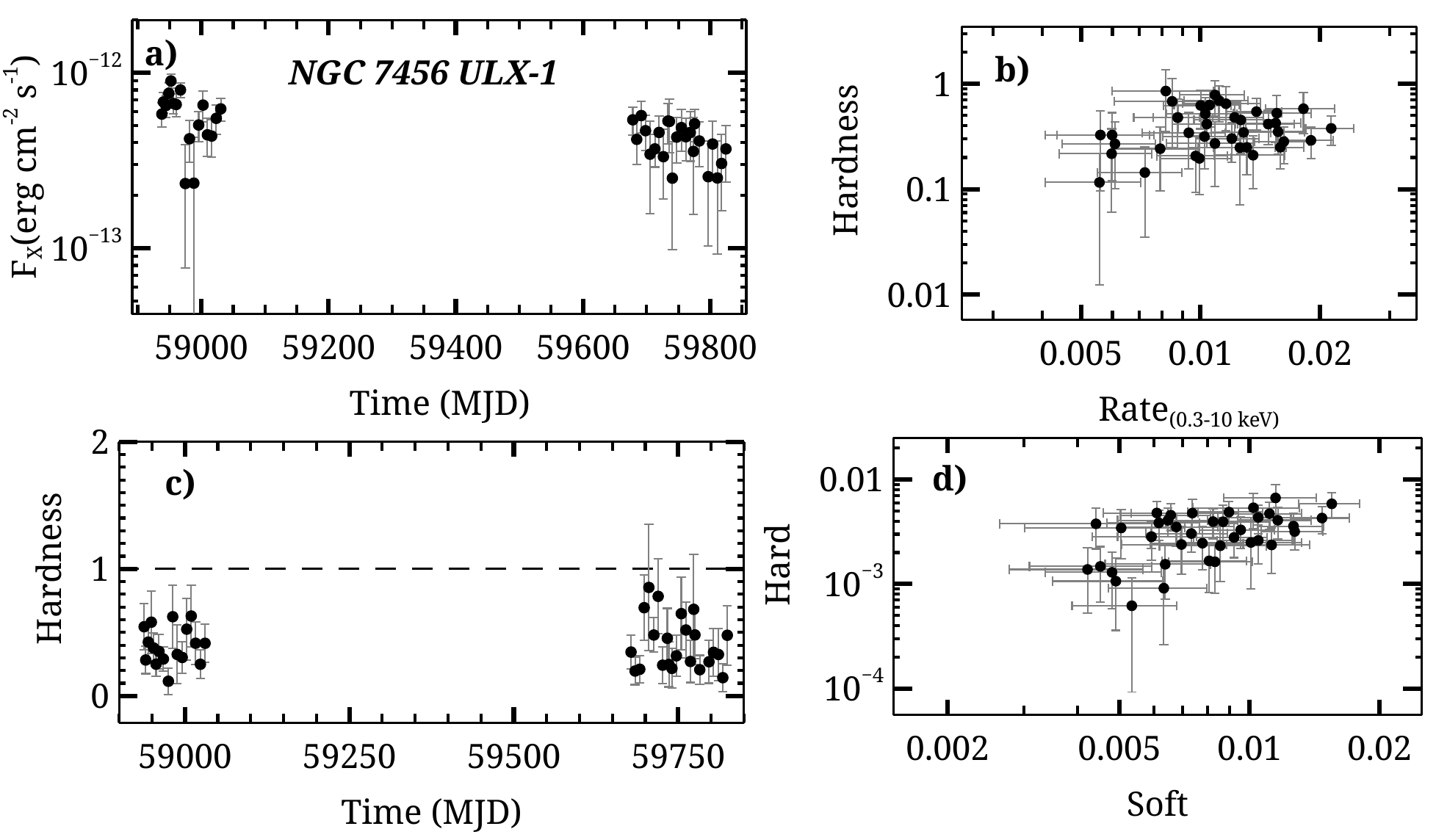}}
\caption{Same as Fig.~\ref{fig:swiftj0243}, but for NGC 7456 ULX--1.}
\label{NGC7456}
\end{figure}

\begin{figure}
\resizebox{\hsize}{!}{\includegraphics{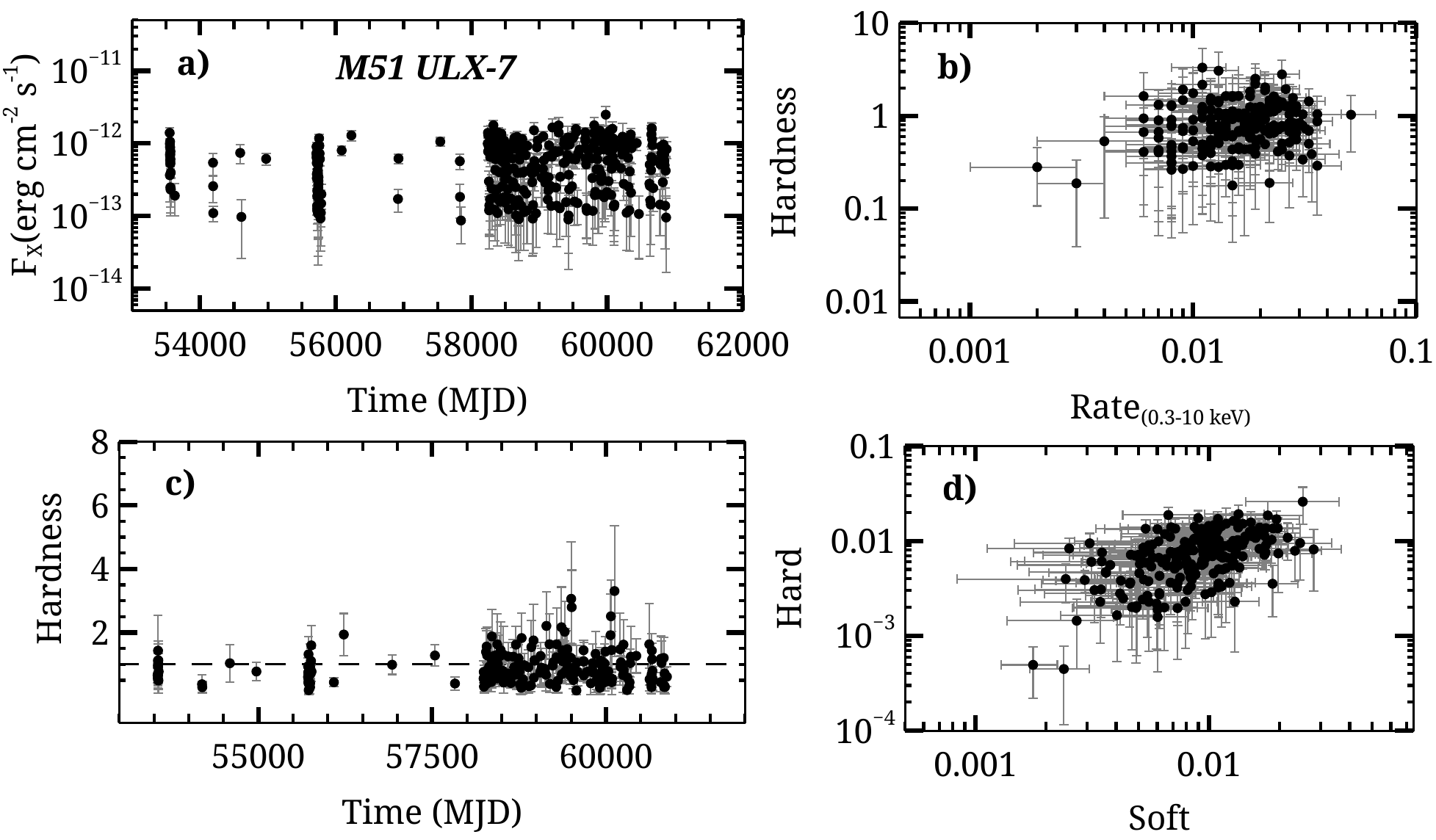}}
\caption{Same as Fig.~\ref{fig:swiftj0243}, but for M51 ULX-7.}
\label{ M51 ULX-7}
\end{figure}

\begin{figure}
\resizebox{\hsize}{!}{\includegraphics{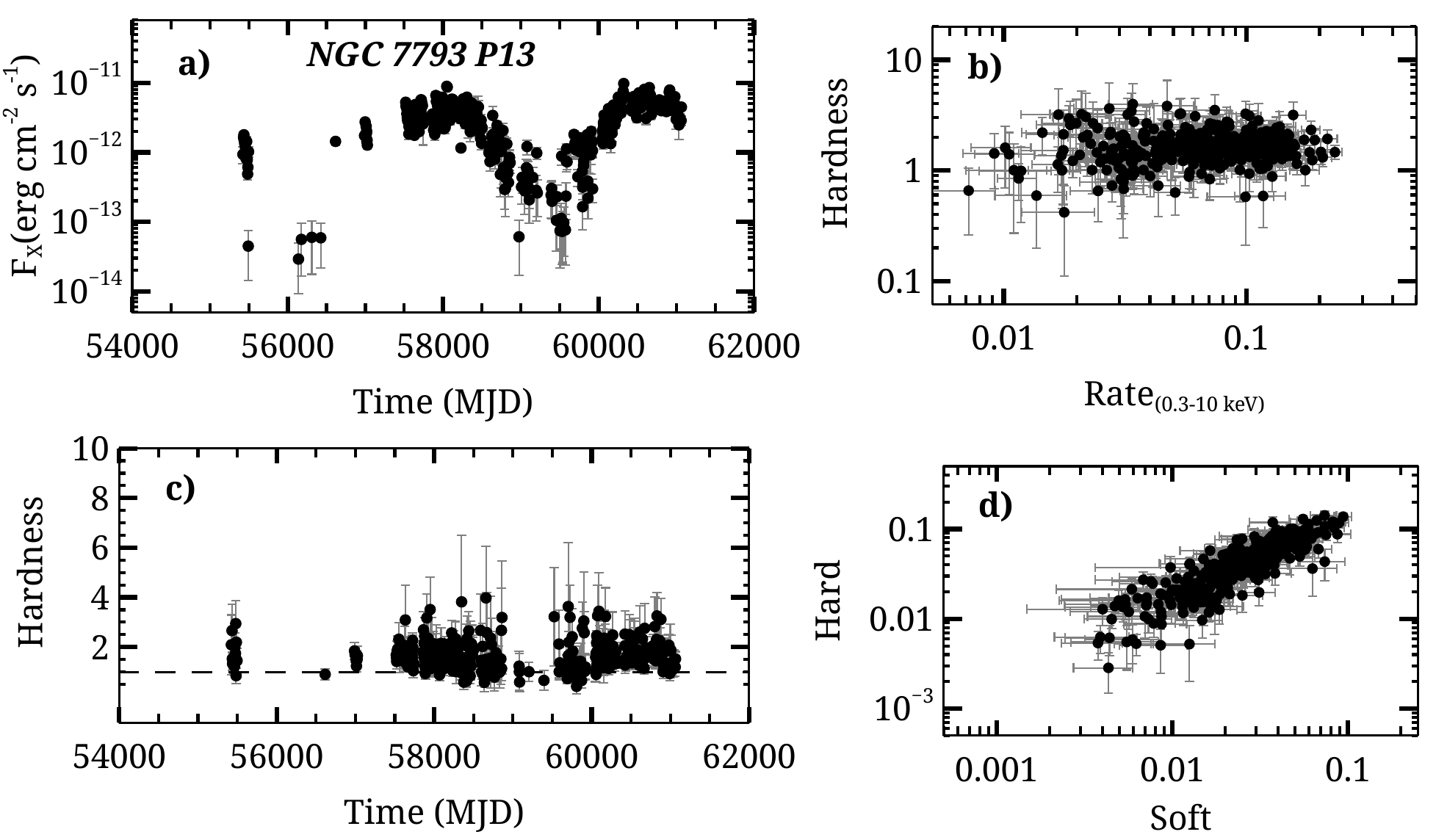}}
\caption{Same as Fig.~\ref{fig:swiftj0243}, but for NGC 7793 P13.}
\label{NGC7793}
\end{figure}

\begin{figure}
\resizebox{\hsize}{!}{\includegraphics{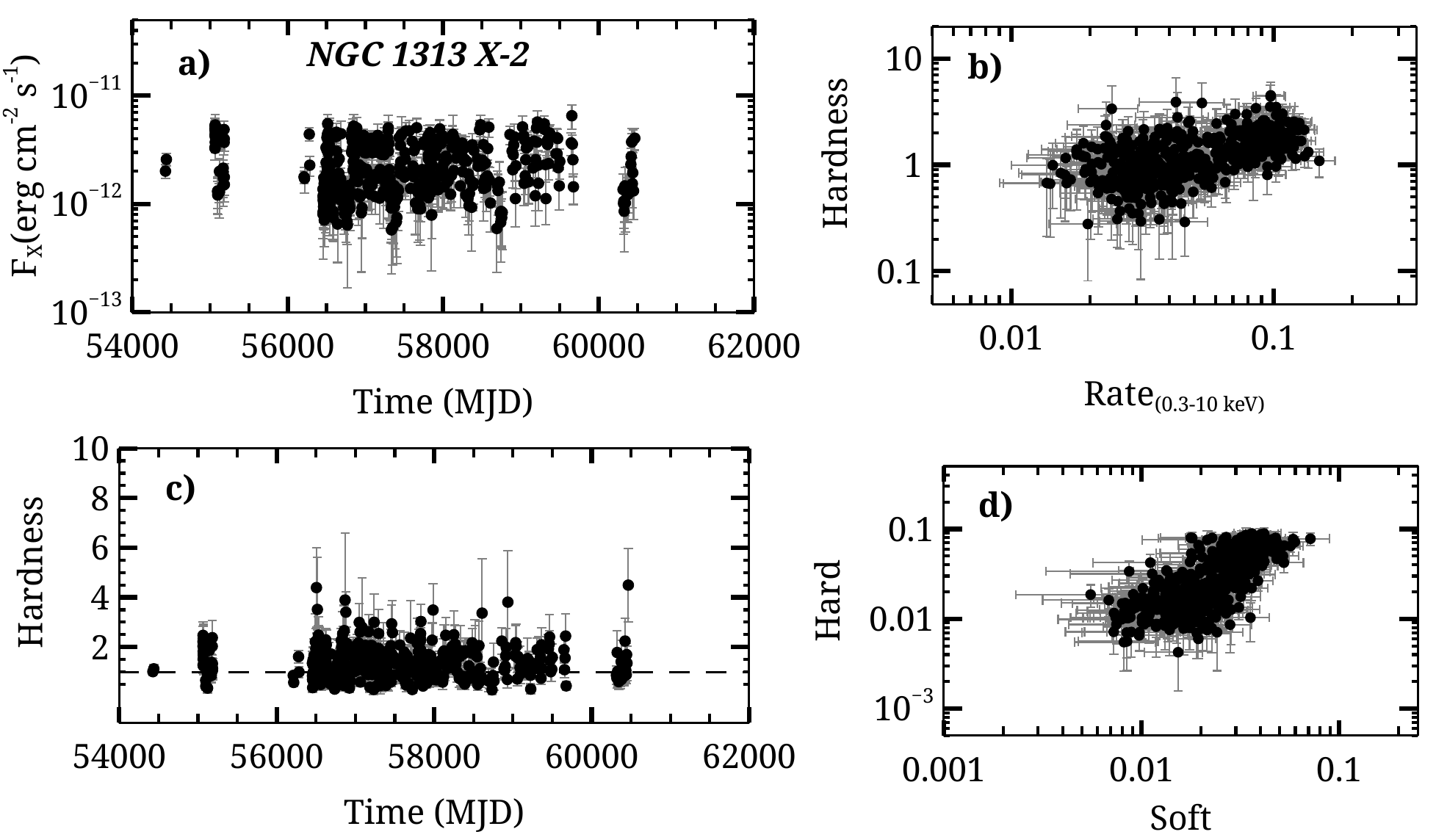}}
\caption{Same as Fig.~\ref{fig:swiftj0243}, but for NGC 1313 X--2.}
\label{NGC1313 X--2}
\end{figure}

\begin{figure}
\resizebox{\hsize}{!}{\includegraphics{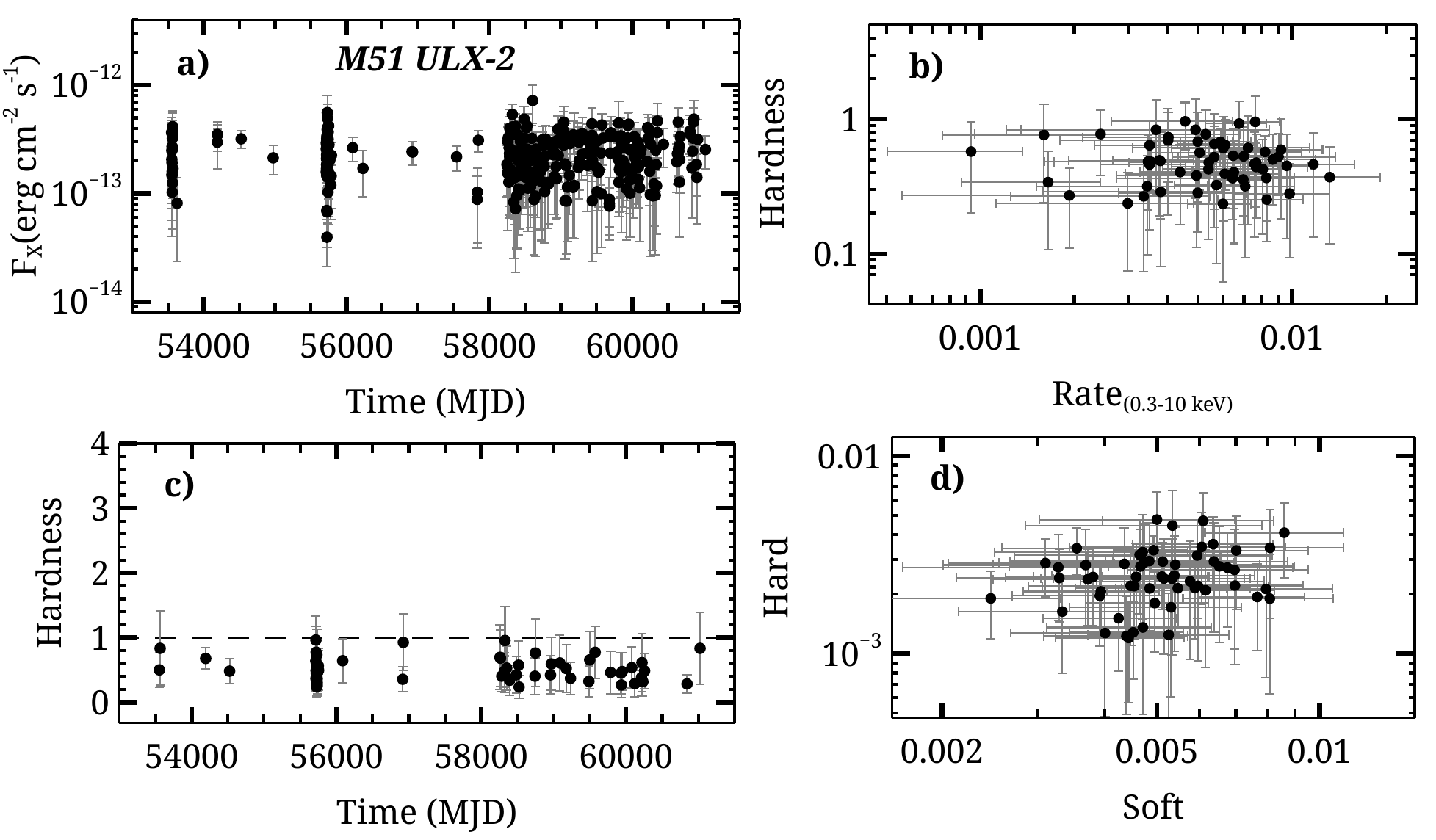}}
\caption{Same as Fig.~\ref{fig:swiftj0243}, but for M51 ULX--2.}
\label{m51ULX--2}
\end{figure}

\begin{figure}
\resizebox{\hsize}{!}{\includegraphics{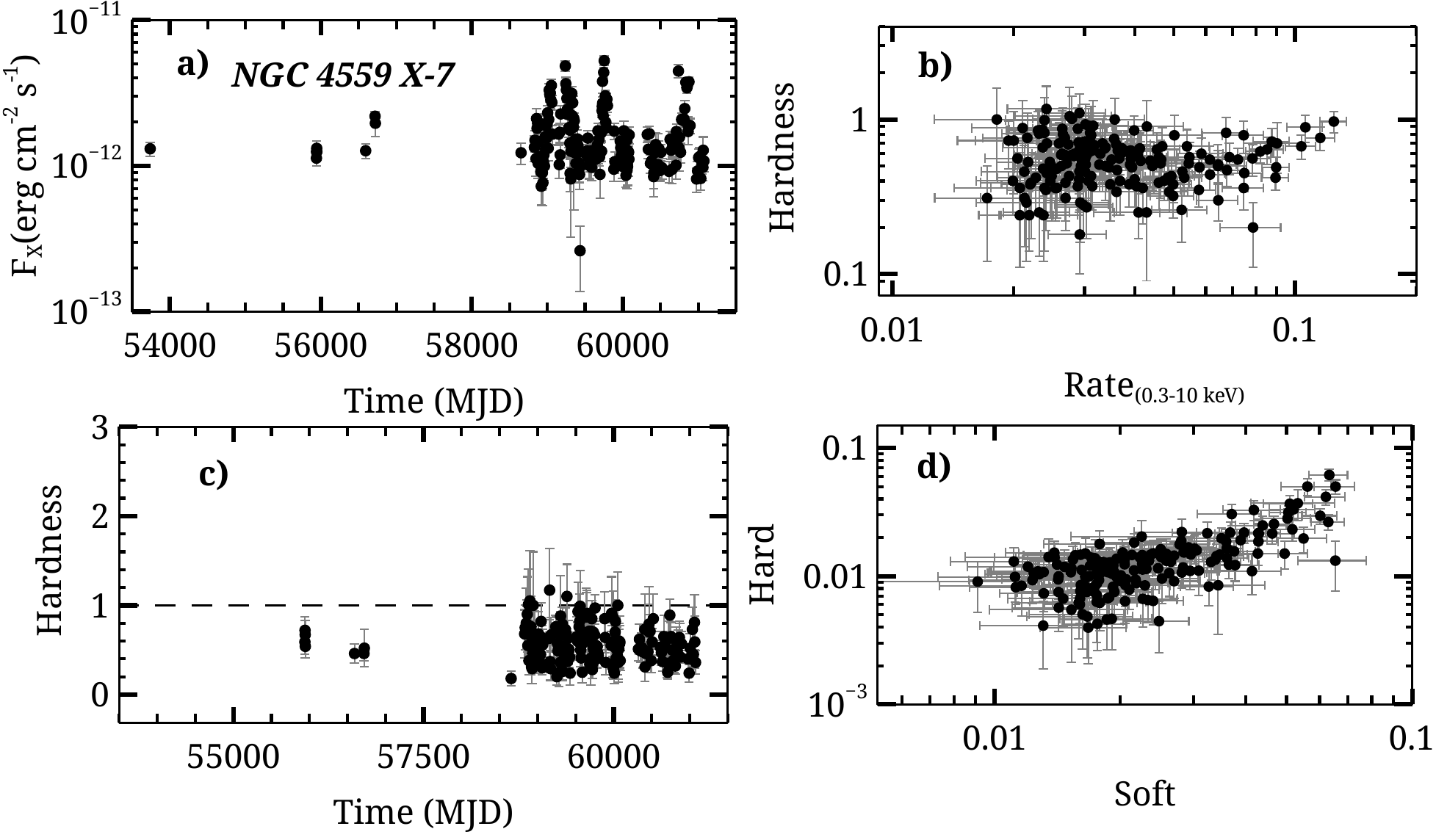}}
\caption{Same as Fig.~\ref{fig:swiftj0243}, but for NGC 4559 X7.}
\label{NGC 4559 X7}
\end{figure}

\begin{figure}
\resizebox{\hsize}{!}{\includegraphics{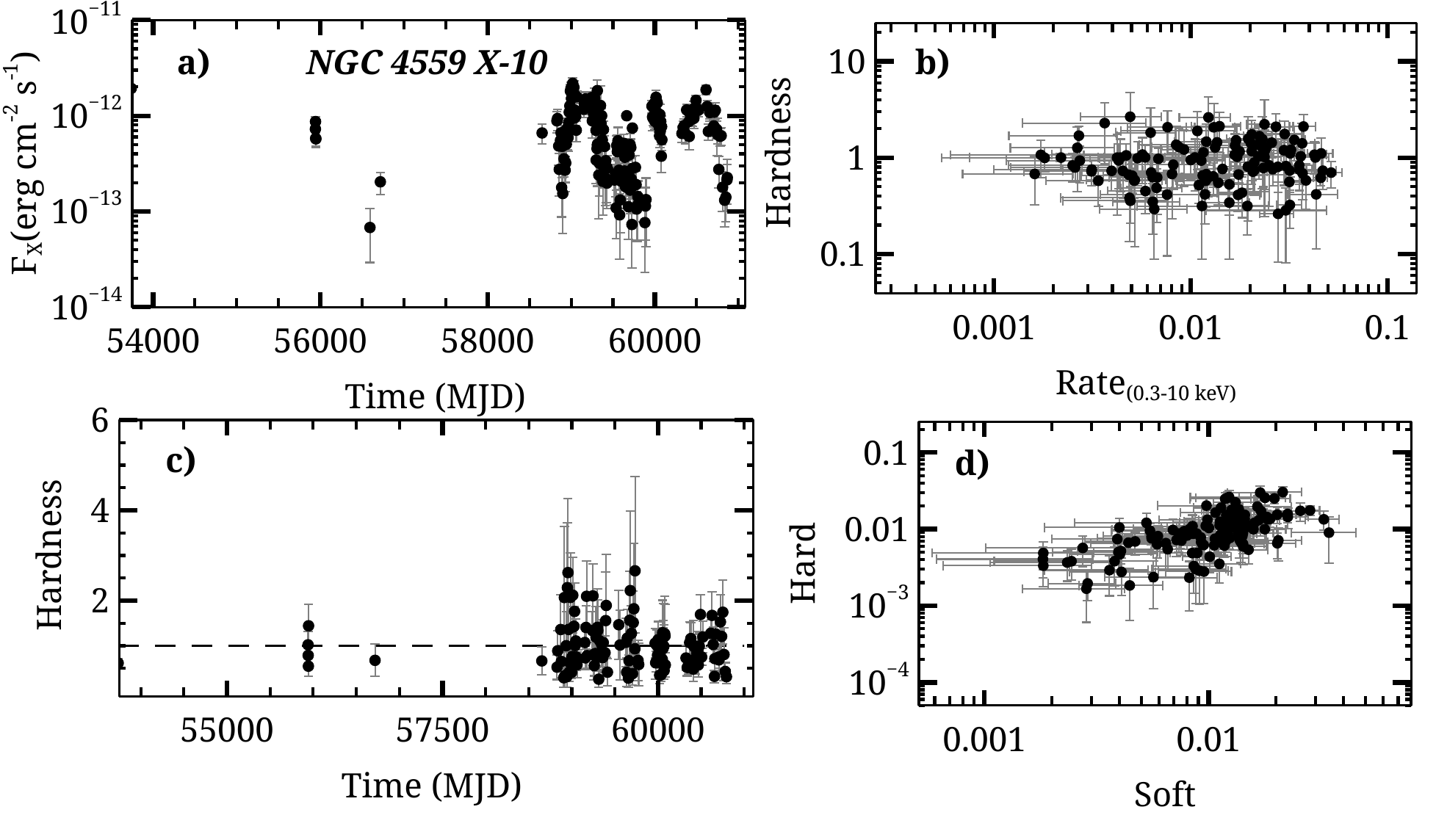}}
\caption{Same as Fig.~\ref{fig:swiftj0243}, but for NGC 4559 X10.}
\label{NGC 4559 X10}
\end{figure}

\begin{figure}
\resizebox{\hsize}{!}{\includegraphics{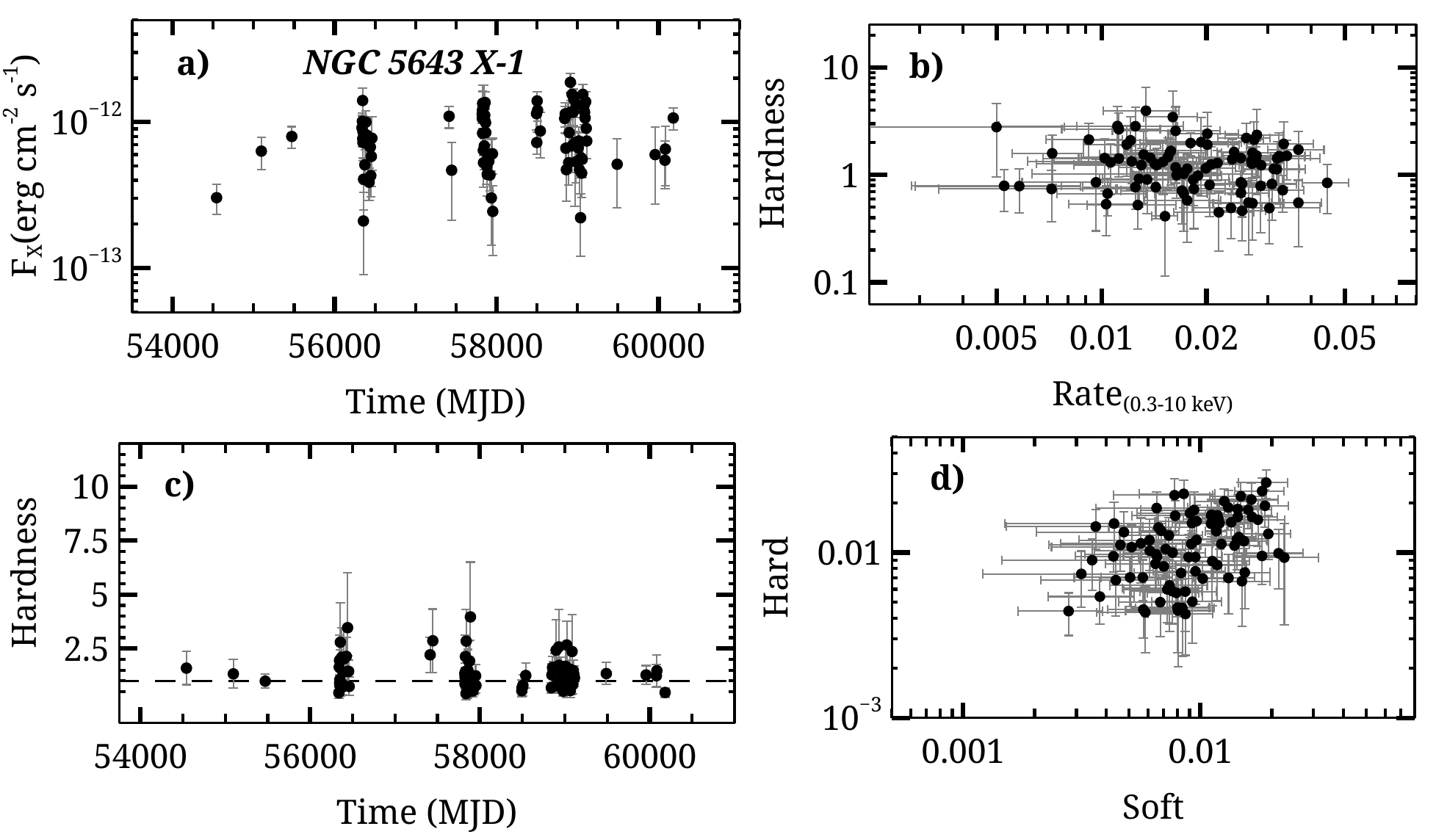}}
\caption{Same as Fig.~\ref{fig:swiftj0243}, but for NGC 5643 X-1.}
\label{NGC 5643 X-1}
\end{figure}

\begin{figure}
\resizebox{\hsize}{!}{\includegraphics{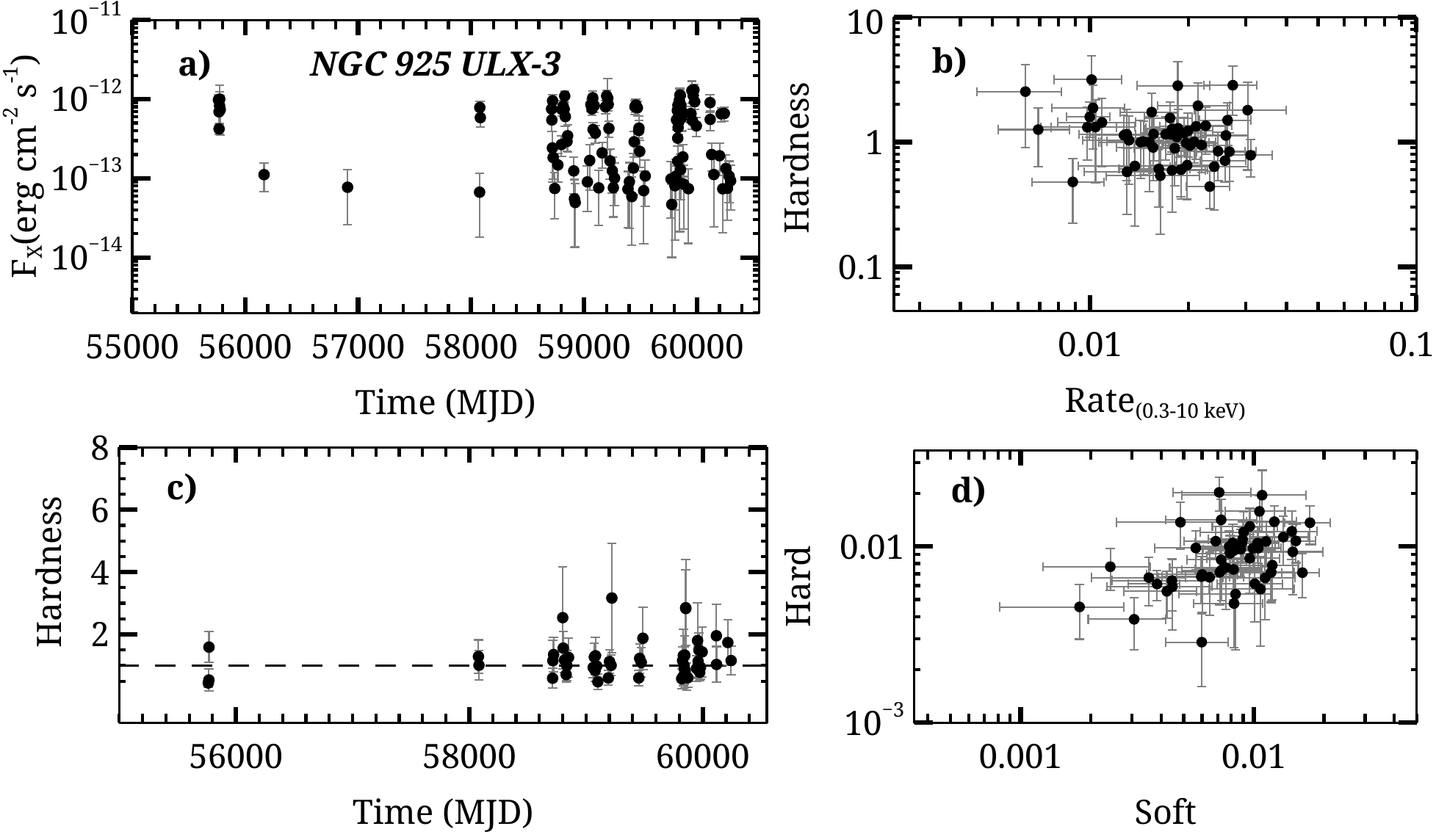}}
\caption{Same as Fig.~\ref{fig:swiftj0243}, but for NGC 925 ULX-3.}
\label{NGC 925 ULX-3}
\end{figure}

\begin{figure}
\resizebox{\hsize}{!}{\includegraphics{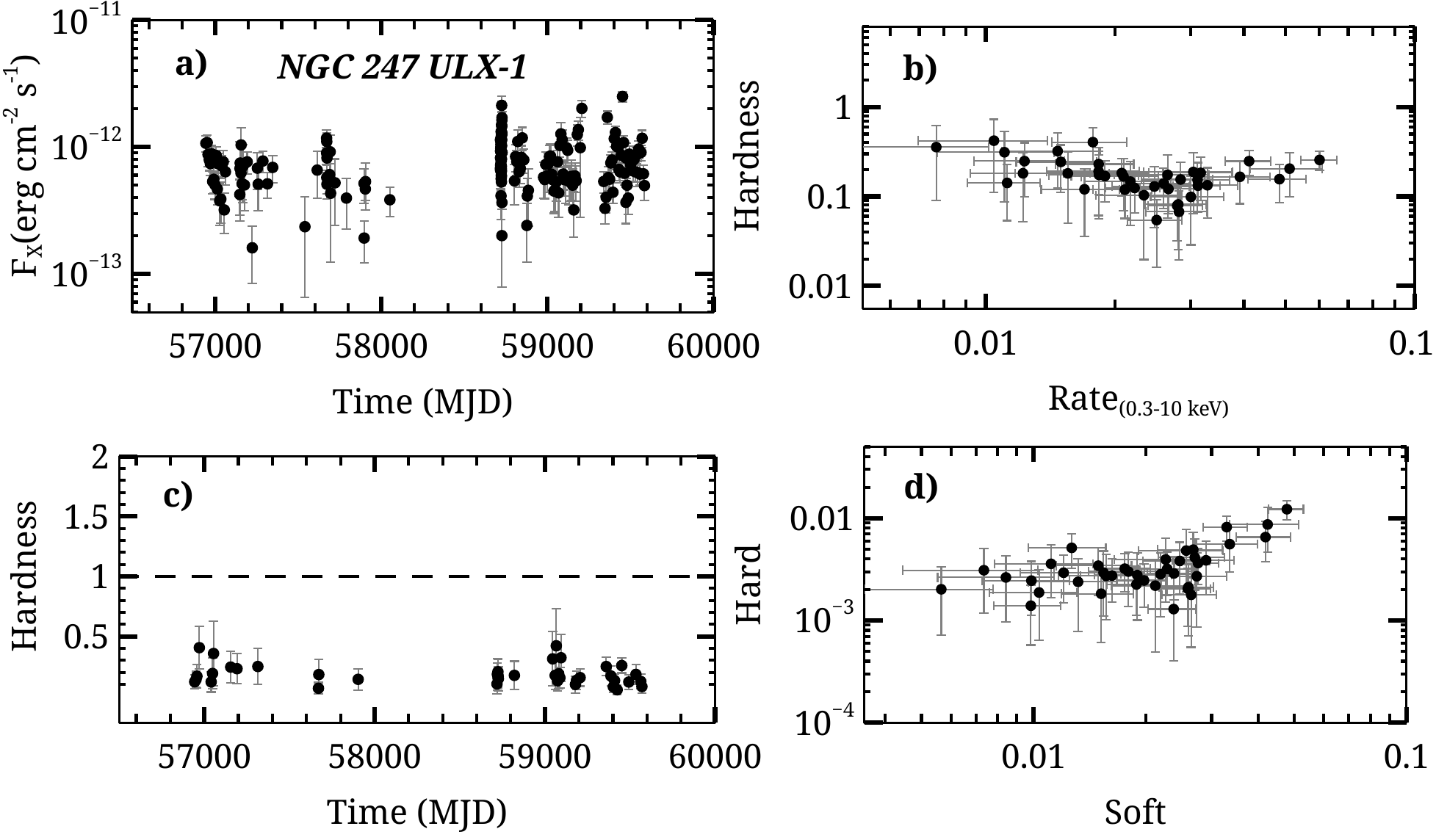}}
\caption{Same as Fig.~\ref{fig:swiftj0243}, but for NGC 247 ULX--1.}
\label{NGC 247 ULX--1}
\end{figure}

\end{appendix}

\end{document}